\documentclass[12pt]{JHEP3}
\usepackage[centertags]{amsmath}
\usepackage{amssymb}
\usepackage{epsfig}
\usepackage{amsmath}
\usepackage{graphicx}
\usepackage{afterpage}

\def\beq{\begin{equation}}
\def\eeq{\end{equation}}
\def\bea{\begin{eqnarray}}
\def\eea{\end{eqnarray}}
\def\nn{\nonumber}
\def\nl{\nonumber\\}

\def\roughly#1{\mathrel{\raise.3ex\hbox
{$#1$\kern-.75em\lower1ex\hbox{$\sim$}}}}

\def\lesssim{\mathrel{\hbox{\rlap{\hbox{\lower4pt\hbox{$\sim$}}}\hbox{$<$}}}}
\def\gtrsim{\mathrel{\hbox{\rlap{\hbox{\lower4pt\hbox{$\sim$}}}\hbox{$>$}}}}

\def\sla#1{\raise.15ex\hbox{$/$}\kern-.57em #1}

\def\bs{B_s^0}
\def\bdbar{{\bar B}_d^0}
\def\bsbar{{\bar B}_s^0}

\def\btos{ b \to  s}
\def\bsmumu{ b \to  s \mu^+ \mu^-}

\def\AFB{A_{FB}}

\def \kstar{{\bar{K}^*}}
\def\Bsmumu{\bsbar \to \mu^+ \mu^-}
\def\Bsmumugamma{\bsbar \to \mu^+ \mu^- \gamma}
\def\BKmumu{\bdbar \to {\bar K} \mu^+ \mu^-}
\def\BKstarmumu{\bdbar\to \kstar \mu^+ \mu^-}
\def\BXsmumu{\bdbar \to X_s \mu^+ \mu^-}

\newcommand{\ba}{\begin{array}}
\newcommand{\ea}{\end{array}}

\def\barr{\begin{eqnarray}}
\def\earr{\end{eqnarray}}
\def\beast{\begin{eqnarray*}}
\def\eeast{\end{eqnarray*}}

\def\be{\begin{equation}}
\def\ee{\end{equation}}
\def\bea{\begin{eqnarray}}
\def\eea{\end{eqnarray}}

\title{\boldmath New Physics in $\bsmumu$: \\
~~~CP-Violating Observables}

\author{
Ashutosh Kumar Alok$^a$,
Alakabha Datta$^b$,
Amol Dighe$^c$,
Murugeswaran Duraisamy$^b$,
Diptimoy Ghosh$^c$, and
David London$^a$
\\
$^a$ Physique des Particules, Universit\'e de Montr\'eal,
\\ ~~C.P. 6128, succ. centre-ville, Montr\'eal, QC, Canada H3C 3J7 \\
$^b$ Department of Physics and Astronomy, 108 Lewis Hall,
\\ ~~University of Mississippi, Oxford, MS 38677-1848, USA \\
$^c$ Tata Institute of Fundamental Research, Homi Bhabha Road,
\\ ~~Mumbai 400005, India \\
E-mail:
\email{alok@lps.umontreal.ca},
\email{datta@phy.olemiss.edu},
\email{amol@theory.tifr.res.in},
\email{duraism@phy.olemiss.edu},
\email{diptimoyghosh@theory.tifr.res.in},
\email{london@lps.umontreal.ca}
}

\abstract{We perform a comprehensive study of the impact of
  new-physics operators with different Lorentz structures on
  CP-violating observables involving the $b \to s \mu^+ \mu^-$
  transition.  We examine the effects of new vector-axial vector (VA),
  scalar-pseudoscalar (SP) and tensor (T) interactions on the CP
  asymmetries in the branching ratios and forward-backward asymmetries
  of $\Bsmumu$, $\BXsmumu$, $\Bsmumugamma$, $\BKmumu$, and
  $\BKstarmumu$.  In $\BKstarmumu$, we also explore the direct CP
  asymmetries in the longitudinal polarization fraction $f_L$ and the
  angular asymmetries $A_T^{(2)}$ and $A_{LT}$, as well as the
  triple-product CP asymmetries $A_T^{(im)}$ and $A^{(im)}_{LT}$.  We
  find that, in almost all cases, the CP-violating observables are
  sensitive only to new physics which involves VA operators. The VA
  new physics may therefore be unambiguously identified by a combined
  analysis of future measurements of these CP-violating observables.
}


\preprint{UdeM-GPP-TH-11-197, TIFR/TH/11-09, UMISS-HEP-2011-02}

\begin{document}

\section{Introduction}

The $B$ factories have taken us to the luminosity frontier with more
than a billion $B^+ / B_d$ mesons, and the Tevatron experiments have
provided us with invaluable data on $B_s$ mesons. We have now entered
the precision era of $B$ physics.  The Standard Model (SM) has been
successful in explaining most of the data to date.  However, this is
now the time to look forward to precision tests, with the ATLAS and
CMS experiments already running, the LHCb expected to start recording
data soon, and the Super-$B$ factories on their way.  One can now be
ambitious and not only look for new-physics (NP) effects, but also try
to identify the kind of NP involved.

Though there is no unambiguous signal of NP so far in all of the $B$
decays we have observed, some possible hints of NP have recently
surfaced in modes involving $b \to s$ transitions.  These include
measurements of CP-averaged quantities such as the large transverse
polarization in $B \to \phi K^*$ \cite{phiK*-babar,phiK*-belle}, and
the anomalous forward-backward asymmetry in $B \to K^* \mu^+ \mu^-$
\cite{Belle-newKstar,BaBar-Kstarmumu,AFBNP}.  There are also
measurements of CP-violating quantities such as the difference between
the mixing-induced CP asymmetries seen in $b \to s$ penguin decays and
in $B_d \to J/\psi K_S$ \cite{btos-1,btos-2,btos-3}, the large CP
asymmetry in $B_s \to J/\psi \phi$ \cite{cdf-d0-note}, and the anomalous CP
asymmetry in like-sign dimuon signals \cite{D0-dimuon}.

In the companion paper \cite{Alok:2010zd}, we performed a general
analysis with all possible Lorentz structures of NP in the transition
$\bsmumu$.  We included NP vector-axial vector (VA),
scalar-pseudoscalar (SP), and tensor (T) $\bsmumu$ operators, and
explored their possible effects on the decays $\Bsmumu$, $\BXsmumu$,
$\Bsmumugamma$, $\BKmumu$, and $\BKstarmumu$.  We focused on
CP-conserving observables such as differential branching ratios,
forward-backward asymmetries, polarization fractions, and the
asymmetries $A_T^{(2)}, A_{LT}$ in $\BKstarmumu$.  Because we only
considered CP-conserving observables, all the
NP couplings were taken to be real.  We computed the effects of all NP
operators, individually and in all combinations, on these observables.

The CP-violating observables in various $\bsmumu$ decays in the SM as
well as in some NP models have been studied in
Refs.~\cite{Du:1995ez,Aliev:1995gm,Fukae:2001dm,Alok:2008dj,Soni:2010xh,Balakireva:2009kn,
  Balakireva:2010zz,Aliev:2005pwa,Kruger:1999xa,Aliev:1999re,Buchalla:2000sk,
  Kruger:2000zg,Bobeth:2008ij,Altmannshofer:2008dz,Egede:2009tp,Egede:2010zc}
In this paper, we explore the CP-violating quantities that may be
measured in the same decay modes by allowing the new couplings to be
complex. The introduction of complex couplings has two effects.
First, some quantities which were taken to be CP-conserving above now
display CP-violation, i.e. the quantities take different values in the
CP-conjugate decays.  The difference between the value of a
measurement in a decay and in its CP-conjugate counterpart is then a
CP-violating observable. Second, new observables appear which vanish
in the CP-conserving limit. (These were not considered in
Ref.~\cite{Alok:2010zd} for this reason.)  These essentially
correspond to the CP-violating triple-product asymmetries $A_T^{(im)}$
and $A_{LT}^{(im)}$ in $\BKstarmumu$, which may be obtained from the
angular distribution in this decay.  Our goal is to identify those
quantities for which there may be large effects due to the presence of
NP. In such cases, we try to find salient features of the effects of
NP, which may help us identify the Lorentz structure of the NP
involved.

Here we have taken the NP to be present only in the effective
$\bsmumu$ operator. While this can, in principle, contribute to CP
violation in $B_d$-$\bar{B}_d$ and $B_s$-$\bar{B}_s$ mixing, it is a
higher-order effect, and hence negligible compared to the SM
contribution.  We therefore neglect mixing-induced (indirect) CP
violation in this work, and focus only on CP violation in the decay.
In the SM, such CP violation is expected to be close to zero in
$\btos$ transitions. A naive estimate indicates that this asymmetry
will be $\sim 10^{-3}$ \cite{Kruger:1999xa,Kruger:2000zg}, but even if
next-to-leading order (NLO) QCD corrections and hadronic uncertainties
are included, it is observed that the CP asymmetry will not exceed 1\%
\cite{Bobeth:2008ij,Altmannshofer:2008dz,Bobeth:2011gi}.  Thus, if a
large CP-violating effect, more than a few percent, is observed in any
of the $\bsmumu$ channels, this will therefore be a clear signature of
NP.  In this paper, we go further and explore the extent to which the
Lorentz structure of NP can be ascertained from the CP-violating
measurements.

The paper is organized as follows.  We begin in Sec.~\ref{bsmumuops}
by describing the effective Hamiltonian with NP operators and new
couplings.  Although the formalism is the same as that used in
Ref.~\cite{Alok:2010zd}, the constraints on the NP couplings are now
more relaxed since the couplings are allowed to be complex. We also
present an overview of the types of CP-violating observables which are
examined. In Sec.~\ref{Bsmumu} we note that there are essentially no
measurable CP-violating quantities in the mode $\Bsmumu$. We then
consider the decays $\BXsmumu$ (Sec.~\ref{BXsmumu}), $\Bsmumugamma$
(Sec.~\ref{Bsmumugamma}), and $\BKmumu$ (Sec.~\ref{BKmumu}). In these
sections we examine the same observables as in
Ref.~\cite{Alok:2010zd}, this time looking at the asymmetries between
these processes and their CP-conjugates.  In Sec.~\ref{BKstarmumu}, we
study the CP asymmetries in $\BKstarmumu$ for the observables
considered in Ref.~\cite{Alok:2010zd}, and in addition we explore new
observables that vanish in the CP-conserving limit (triple products).
We summarize our findings in Sec.~\ref{summary} and discuss their
implications.
 
\section{\boldmath $\bsmumu$ Operators
\label{bsmumuops}}

\subsection{Effective Hamiltonian in the SM and with NP}

Our formalism is identical to that used in Ref.~\cite{Alok:2010zd}.
We repeat it here briefly for the sake of completeness.  Within the
SM, the effective Hamiltonian for the quark-level transition $\bsmumu$
is
\bea
{\cal H}_{\rm eff}^{SM} &=& -\frac{4 G_F}{\sqrt{2}}
\, V_{ts}^* V_{tb} \, \Bigl\{ \sum_{i=1}^{6} {C}_i (\mu) {\cal O}_i (\mu)
+ C_7 \,\frac{e}{16 \pi^2}\, [\bar{s} 
  \sigma_{\mu\nu} (m_s P_L + m_b P_R) b] \,
F^{\mu \nu} \nn \\
&& +\, C_9 \,\frac{\alpha_{em}}{4 \pi}\, (\bar{s}
\gamma^\mu P_L b) \, \bar{\mu} \gamma_\mu \mu + C_{10}
\,\frac{\alpha_{em}}{4 \pi}\, (\bar{s} \gamma^\mu P_L b) \, \bar{\mu} 
\gamma_\mu \gamma_5  
\mu  \, \Bigr\} + h.c.,
\label{HSM}
\eea
where $P_{L,R} = (1 \mp \gamma_5)/2$. The operators ${\cal O}_i$
($i=1,..6$) correspond to the $P_i$ in Ref.~\cite{bmu}, and $m_b =
m_b(\mu)$ is the running $b$-quark mass in the $\overline{\rm MS}$
scheme.  We use the SM Wilson coefficients ($C_i$) as given in
Ref.~\cite{Altmannshofer:2008dz}.

The effective Hamiltonian in the presence of NP is
\beq
{\cal H}_{\rm eff}(\bsmumu) = {\cal
H}_{\rm eff}^{SM} + {\cal H}_{\rm eff}^{VA} + {\cal H}_{\rm eff}^{SP} +
{\cal H}_{\rm eff}^{T} + h.c.,
\label{NP:effHam}
\eeq
where 
\bea
{\cal H}_{\rm eff}^{VA} &=& - \frac{4 G_F}{\sqrt{2}} \,
\frac{\alpha_{em}}{4\pi} \, V_{ts}^* V_{tb} \, 
\Bigl\{ R_V \, (\bar{s} \gamma^\mu P_L b)
\, \bar{\mu} \gamma_\mu \mu + R_A \, (\bar{s} \gamma^\mu P_L b)
\, \bar{\mu} \gamma_\mu \gamma_5 \mu \nn \\
&& \hskip3.0 truecm +~R'_V \, (\bar{s} \gamma^\mu P_R b) \,
\bar{\mu} \gamma_\mu \mu + R'_A \, (\bar{s} \gamma^\mu P_R b)
\, \bar{\mu} \gamma_\mu \gamma_5\mu \Bigr\} ~, \\
{\cal H}_{\rm eff}^{SP} &=& - \frac{4G_F}{\sqrt{2}} \,
\frac{\alpha_{em}}{4\pi}\, V_{ts}^* V_{tb} \, 
\Bigl\{R_S ~(\bar{s} P_R b) ~\bar{\mu}\mu +
R_P ~(\bar{s} P_R b) ~ \bar{\mu}\gamma_5 \mu \nn\\
&& \hskip3.0 truecm +~R'_S ~(\bar{s} P_L b) ~\bar{\mu}\mu +
R'_P ~(\bar{s} P_L b) ~ \bar{\mu}\gamma_5 \mu \Bigr\} \;, \\
{\cal H}_{\rm eff}^{T} &=& -\frac{4 G_F}{\sqrt{2}} \,
\frac{\alpha_{em}}{4\pi}\, V_{ts}^* V_{tb} \, 
\Bigl\{C_T (\bar{s} \sigma_{\mu \nu } b)
\bar{\mu} \sigma^{\mu\nu}\mu + i C_{TE} (\bar{s} \sigma_{\mu \nu } b) 
\bar{\mu} \sigma_{\alpha \beta } \mu ~\epsilon^{\mu
\nu \alpha \beta} \Bigr\}
\eea
are the new contributions.  Here, $R_V, R_{A}, R_V', R_A', R_S, R_P,
R_S', R_P', C_{T}$ and $C_{TE}$ are the NP effective couplings.  In
our numerical analysis in this paper, we take all NP couplings to be
complex. As in Ref.~\cite{Alok:2010zd}, we do not include NP through
the $O_7 =\bar s\sigma^{\alpha\beta} P_R b \, F_{\alpha\beta}$
operator or its chirally-flipped counterpart $O_7^\prime= \bar
s\sigma^{\alpha\beta} P_L b \,F_{\alpha\beta}$.

\subsection{Constraints on NP couplings}
\label{constraints}

The constraints on the NP couplings in $\bsmumu$ come mainly from the
upper bound on the branching ratio $B(\Bsmumu)$ and the measurements
of the total branching ratios $B(\BXsmumu)$ and $B(\BKmumu)$
\cite{pdg,Barberio:2008fa,Huber:2007vv}:
\barr
B(\Bsmumu) & < & 4.70 \times 10^{-8} \quad \mbox{(90\% C.L.)} \; , \\
B(\BXsmumu) & = & \left\{ \begin{array}{ll}
\left( 1.60 \pm 0.50 \right) \times 10^{-6} & (\mbox{low } q^2)  \\
\left( 0.44 \pm 0.12 \right) \times 10^{-6} & (\mbox{high } q^2)  \\   
\end{array} \right. \; , \\
B(\BKmumu) & = & \left(4.5^{+1.2}_{-1.0} \right) \times 10^{-7} \; ,
\earr
where the low-$q^2$ and high-$q^2$ regions correspond to 1 GeV$^2 \le 
q^2 \le  6$ GeV$^2$ and $q^2 \ge 14.4$ GeV$^2$, respectively. 
Here $q^2$ is the invariant mass squared of the two muons. 

We consider all the NP couplings $R_i$ to be complex and parametrize
them as 
\begin{equation}
R_{i} = |R_i| \, e^{i \phi_{R_i}}\,,
\end{equation}
where $i=V,A,S,P,T,TE$ and $-\pi \leq \phi_{R_i} \leq \pi$.
The bounds on these couplings will in general depend on which operators
are present. While we take the correlations in these constraints into
account in our numerical calculations, for the sake of simplicity we
only give the bounds when the NP operators (VA, SP, T) are present
individually.

If the only NP couplings present are $R_{V,A}$, we obtain
\beq
\frac{|{\rm Re}(R_V) + 2.8 |^2}{(6.3)^2} + 
\frac{|{\rm Im}(R_V)|^2}{(6.0)^2} 
\lesssim 1.0 \; , \quad
\frac{|{\rm Re}(R_A) -4.1 |^2}{(6.1)^2} + 
\frac{|{\rm Im}(R_A)|^2}{(6.0)^2} 
\lesssim 1.0  \; .
\eeq
If the only NP couplings present are $R'_{V,A}$, the constraints are
\beq
\frac{|{\rm Re}(R'_V)|^2}{(3.5)^2} + 
\frac{|{\rm Im}(R'_V)|^2}{(4.0)^2} 
\lesssim 1.0 \; , \quad
\frac{|{\rm Re}(R'_A)|^2}{(3.5)^2} + 
\frac{|{\rm Im}(R'_A)|^2}{(4.0)^2} 
\lesssim 1.0 \; .
\eeq
For the SP operators, the present upper bound on $B(\Bsmumu)$ provides
the limit
\begin{equation}
|R_S - R'_S|^2 + |R_P - R'_P|^2 \lesssim 0.44 ~.
\label{SP-constraints}
\end{equation}
This constitutes a severe constraint on the NP couplings if only
$R_{S,P}$ or $R'_{S,P}$ are present. However, if both types of
operators are present, these bounds can be evaded due to cancellations
between the $R_{S,P}$ and $R'_{S,P}$.  In that case, $B(\BXsmumu)$ and
$B(\BKmumu)$ can still bound these couplings. The stronger bound is
obtained from the measurement of the latter quantity, which yields
\beq
|R_S|^2 + |R_P|^2 \lesssim 9 \; ,
\quad R_S \approx R'_S \; , \quad R_P \approx R'_P \; .
\eeq
Finally, the constraints on the NP tensor operators come entirely from
$B(\BXsmumu)$.  When only the T operators are present,
\begin{equation}
|C_T|^2 +4 |C_{TE}|^2 \lesssim 1.0 ~.
\end{equation}
The constraints are not affected significantly if more than one type
(VA, SP or T) of NP operators is present simultabeously.

\subsection{CP-violating effects}
\label{cp-sm-np}

All CP-violating effects are due to the interference of (at least) two
amplitudes with a relative weak phase. In principle, there can be
three types of interference: SM-SM, SM-NP, NP-NP. In the SM, all
contributions to the $\bsmumu$ modes are proportional to the
Cabibbo-Kobayashi-Maskawa (CKM) factors $V_{tb}^*V_{ts}$,
$V_{cb}^*V_{cs}$, or $V_{ub}^*V_{us}$. The term $V_{cb}^*V_{cs}$ can
be eliminated in terms of the other two using the unitarity of the CKM
matrix. Furthermore, although $V_{ub}^*V_{us}$ has a large weak phase,
its magnitude is greatly suppressed relative to that of
$V_{tb}^*V_{ts}$. Thus, to a good approximation, all nonzero SM
contributions have the same weak phase, and so all CP-violating
effects are predicted to be tiny in the SM.

There are two types of CP violation.  The first is direct CP-violating
asymmetries.  Suppose that a particular ${\bar B}$ decay has two
contributing amplitudes: $i{\cal M}({\hbox{${\bar B}$ decay}}) = {\cal
  A}_1 + {\cal A}_2$. Each amplitude has both a weak and a strong
phase. The matrix element $i\overline{\cal M}$ for the CP-conjugate
decay is the same as $i{\cal M}$, except that the weak phases change
signs. CP violation is indicated by a nonzero value of $|{\cal M}|^2 -
|\overline{\cal M}|^2$. It is straightforward to show that this is
proportional to $\sin{\phi_w}\sin{\delta}$, where $\phi_w$ and
$\delta$ are, respectively, the relative weak and strong phases
between ${\cal A}_1$ and ${\cal A}_2$. Direct CP-violating asymmetries
therefore require that the interfering amplitudes have both a nonzero
relative weak and strong phase.

The second type of CP violation is triple-product (TP) asymmetries.
Suppose that the matrix element for the ${\bar B}$ decay takes the
form $i{\cal M}({\hbox{${\bar B}$ decay}}) = {\cal A}_1 + i {\cal A}_2
\epsilon_{\mu\nu\rho\sigma} p_{\bar B}^\mu v_1^\nu v_2^{\rho}
v_3^{\sigma}$, where the $v_i$ are spins or momenta of the final-state
particles. The difference $|{\cal M}|^2 - |\overline{\cal M}|^2$ is
proportional to $m_{\bar B} \vec v_1 \cdot (\vec v_2 \times \vec v_3)
\sin{\phi_w}\cos{\delta}$. By measuring the TP $\vec v_1 \cdot (\vec
v_2 \times \vec v_3)$ in both ${\bar B}$ and $B$ decays, the TP
asymmetry can be obtained.  Note that the measurement of a nonzero TP
in the $\bar{B}$ decay alone is not sufficient to establish CP
violation, i.e.\ it does not necessarily imply a nonzero weak phase.
A fake, CP-conserving TP can be produced if ${\cal A}_1$ and ${\cal
  A}_2$ have a relative strong phase. It is only by measuring the
difference of TPs in ${\bar B}$ and $B$ decays that the fake TP can be
eliminated and a true, CP-violating signal produced \cite{fake-TP}.

Let us first turn to direct CP violation, which requires both a
relative weak and strong phase between two interfering amplitudes.
Now, strong phases are generated through the rescattering of the
operators in the effective Hamiltonian. The NP strong phases involve
only the (constrained) NP operators, and are therefore small
\cite{Datta:2004re}. Thus, direct CP asymmetries can never arise from
NP-NP interference.

On the other hand, the SM strong phase is not so small. It is
generated because the Wilson coefficient $C^{\rm eff}_9$, which gets a
contribution from a $c\bar{c}$ quark loop, has an imaginary
piece. ($C^{\rm eff}_9$ also gets a contribution from a $u\bar{u}$
quark loop. But this is proportional to $V_{ub}^*V_{us}$, and hence
negligible.) The quantity $C^{\rm eff}_9$ can be written as
\cite{Altmannshofer:2008dz}
\begin{eqnarray}
\label{effecWC1}
C^{\rm eff}_9 &\!=\!& C_9(m_b) + h(z,\hat{m_c}) \left(\frac{4}{3} C_1 + C_2 + 6\, C_3 + 60\, 
C_5 \right) \nn\\
&&-~\frac{1}{2} h(z,\hat{m_b}) \left(7 C_3 + \frac{4}{3} C_4 + \, 76 
C_5 + 
\frac{64}{3} C_6 \right) \\
&&-~\frac{1}{2} h(z,0) \left( C_3 + \frac{4}{3} C_4 + 16\, C_5 + 
\frac{64}{3} C_6 
\right) + \frac{4}{3} C_3 + \frac{64}{9} C_5 + \frac{64}{27} C_6 ~. \nn
\end{eqnarray}

Here $z \equiv q^2/m_b^2$, and $\hat{m}_q \equiv m_q/m_b$ for all
quarks $q$.  The function $h(z,\hat m)$ represents the one-loop
correction to the four-quark operators $O_1$-$O_6$ and is given by
\cite{Buras:1994dj,Kruger:2000zg,Altmannshofer:2008dz}
 \begin{eqnarray}
 \label{effecWC}
 h(z,\hat m) &  = & -\frac{8}{9}\ln\frac{m_b}{\mu_b} - \frac{8}{9}\ln \hat m +
 \frac{8}{27} + \frac{4}{9} x \\
 & & - \frac{2}{9} (2+x) |1-x|^{1/2} 
 \left\{\begin{array}{ll}
 \left( \ln\left| \frac{\sqrt{1-x} + 1}{\sqrt{1-x} - 1}\right| - i\pi \right), &
 \mbox{for } x \leq 1 ~, \nonumber \\
 2 \arctan \frac{1}{\sqrt{x-1}}, & \mbox{for } x > 1 ~,
 \end{array}
 \right.\ 
 \end{eqnarray}
where $x \equiv {4\hat m^2}/{z}$.  Thus, a nontrivial strong phase is
generated when $z \geq 4 \hat{m}^2$.  This leads to the complex nature
of $C^{\rm eff}_9$ in the SM. For example, typical values of $C^{\rm
  eff}_9$ in the low- and high-$q^2$ regions are $C^{\rm eff}_9(m_b) =
4.75 + 0.09 i \; (z = 0.1) \; , C^{\rm eff}_9(m_b) = 4.76 + 0.88 i \;
(z = 0.7)$.  $C^{\rm eff}_9$ therefore has a nontrivial imaginary
component, which implies that direct CP asymmetries can arise due to
SM-NP interference.  Since the SM operator ($C^{\rm eff}_9$) is of VA
type, the NP operator must also be VA in order to generate a
significant direct CP asymmetry.  Other NP operators can also
interfere with the SM, but the effect is suppressed by $m_\mu/m_b$,
and hence very small.  Note that, although this argument has used the
total decay rate for illustration, we could have used (almost) any
observable which is related to the square of the matrix element. This
includes the differential branching ratio, forward-backward asymmetry,
polarization asymmetries, etc.

The TP asymmetries, on the other hand, do not need a difference in
strong phases between two amplitudes. Indeed, they are proportional to
$\cos\delta$, though they do require a weak-phase difference.  This
means that a TP asymmetry can be produced by either SM-NP or NP-NP
interference. Given that all SM operators are of VA type, the NP must
also be VA if SM-NP interference is the reason for the TP. On the
other hand, if NP-NP interference is involved, this can arise due to
new SP and T operators (other NP-NP interference are possible, but the
effects are suppressed by $m_\mu/m_b$).

In this paper, we explore both sources of CP asymmetries, direct CP
violation and TPs.  While we have checked the effects of SP and T NP
operators on all the observables, we find them to be insignificant in
most places (as expected from the arguments above), and we will
mention them only during the discussion of TP asymmetries, where, in
principle, they may play a significant role.

\section{\boldmath $\Bsmumu$
\label{Bsmumu}}

We begin by considering the direct CP asymmetry in $\Bsmumu$. Helicity
conservation in the decay of $B_s$ or $\bar{B}_s$ implies that the
only final states can be $\mu^+_L \mu^-_L$ or $\mu^+_R \mu^-_R$, which
are $CP$ conjugates.  The only CP-violating observables that can be
constructed are then
\bea
A_{CP}^{RL}(t) & \equiv & 
\frac{ B(\bsbar(t) \to \mu^+_R \mu^-_R) - B(\bs(t) \to \mu^+_L \mu^-_L)}
{ B(\bsbar(t) \to \mu^+_R \mu^-_R) + B(\bs(t) \to \mu^+_L \mu^-_L)}\; ,
\nonumber \\
A_{CP}^{LR}(t) & \equiv &  
\frac{B(\bsbar(t) \to \mu^+_L \mu^-_L) - B(\bs(t) \to \mu^+_R \mu^-_R) }
{B(\bsbar(t) \to \mu^+_L \mu^-_L) + B(\bs(t) \to \mu^+_R \mu^-_R)}\;.
\eea
The CP asymmetry in the longitudinal polarization fraction $A_{LP}$
may be written in terms of these two CP asymmetries.
The measurement of either of these CP asymmetries requires the
measurement of muon polarization, which will be an impossible task for
the upcoming experiments \cite{Alok:2010zd}.  And even if this were
doable, the lack of any sources for different strong phases in the two
CP-conjugate final states implies that the direct CP asymmetry would
vanish even with NP.
We therefore do not study CP violation in $\Bsmumu$.

\section{\boldmath $\BXsmumu$
\label{BXsmumu}}

A model-independent analysis of the CP asymmetry in the differential
branching ratio (DBR) of $\BXsmumu$ was previously carried out in
Ref.~\cite{Fukae:2001dm}.  There, the CP asymmetry in the DBR was
predicted for some specific values of the NP couplings.  However, no
experimental constraints on the parameters were used.  In this paper
we study the CP asymmetry in the DBR, taking into account the
constraints from the present measurements of other related
observables. Moreover, in addition to the CP asymmetry in the DBR, we
also study the CP asymmetry in the forward-backward asymmetry.

The CP asymmetry in DBR of $\BXsmumu$ is defined as
\begin{equation}
A_{\rm CP}(q^2)=\frac{(dB/dz) - (d\overline{B}/dz)} {(dB/dz) 
+ (d\overline{B}/dz)}\;,
\label{Acp-Xsmumu}
\end{equation}
where $z \equiv q^2/m_{b}^2$, and $dB/dz$ and $d\overline{B}/dz$ are
the DBRs of $\BXsmumu$ and its CP-conjugate $B_d^0 \to X_s \mu^+
\mu^-$, respectively. The expression for $(dB/dz)$ has been given in
Ref.~\cite{Alok:2010zd}.

The CP asymmetry in the forward-backward asymmetry $A_{FB}$ is defined
as
\be
\Delta A_{FB}(q^2) \equiv  A_{FB}(q^2) - \overline{A}_{FB}(q^2) \; ,
\label{AFB-Xsmumu}
\ee
where the definition of $A_{FB}$ is given in Ref.~\cite{Alok:2010zd},
and $\overline{A}_{FB}$ is the analogous quantity for the CP-conjugate
decay. Note that while the relevant angle $\theta$ in $\BXsmumu$ is
defined relative to the direction of $\mu^+$, for the CP-conjugate
decay one should define $\theta$ in relation to the direction of
$\mu^-$, and similarly for $A_{FB}$ in other $\bsmumu$ decay modes
below. 

\FIGURE[t]{
\includegraphics[width=0.4\linewidth]{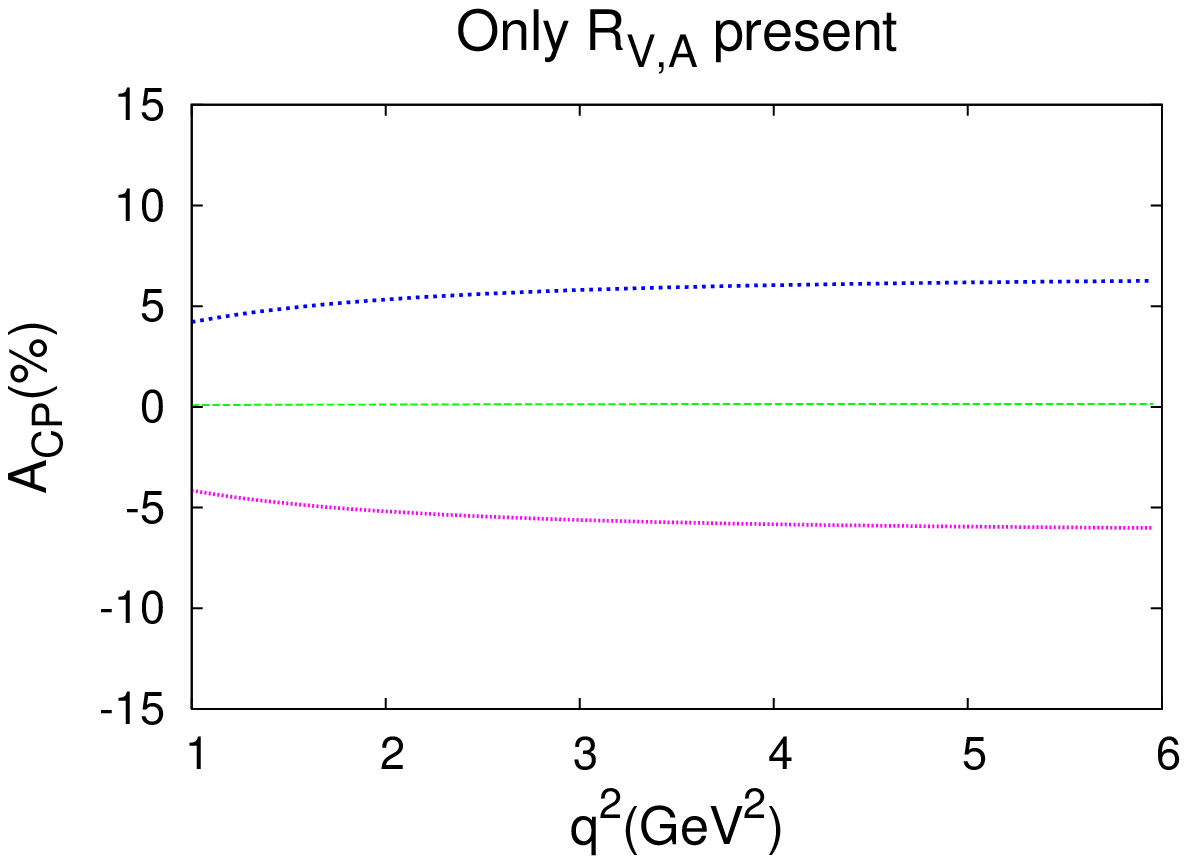} 
\includegraphics[width=0.4\linewidth]{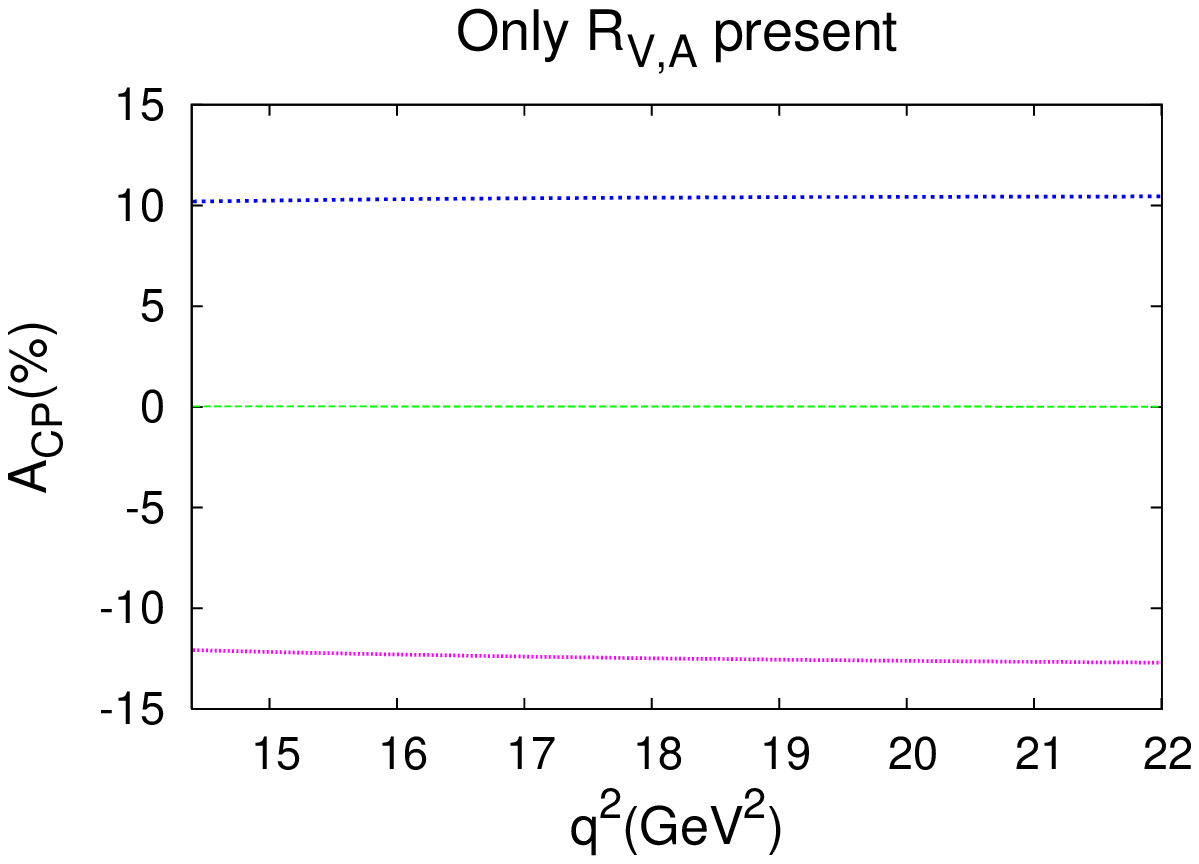}
\includegraphics[width=0.4\linewidth]{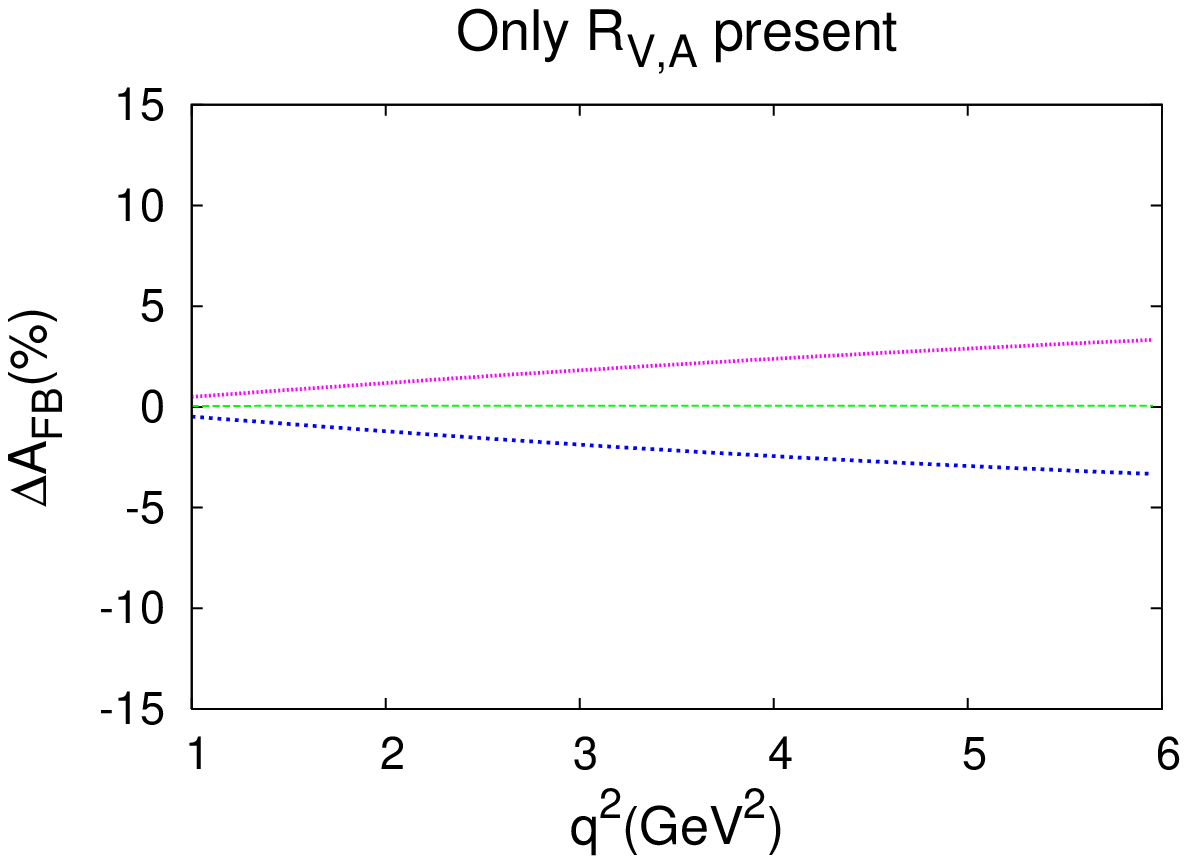} 
\includegraphics[width=0.4\linewidth]{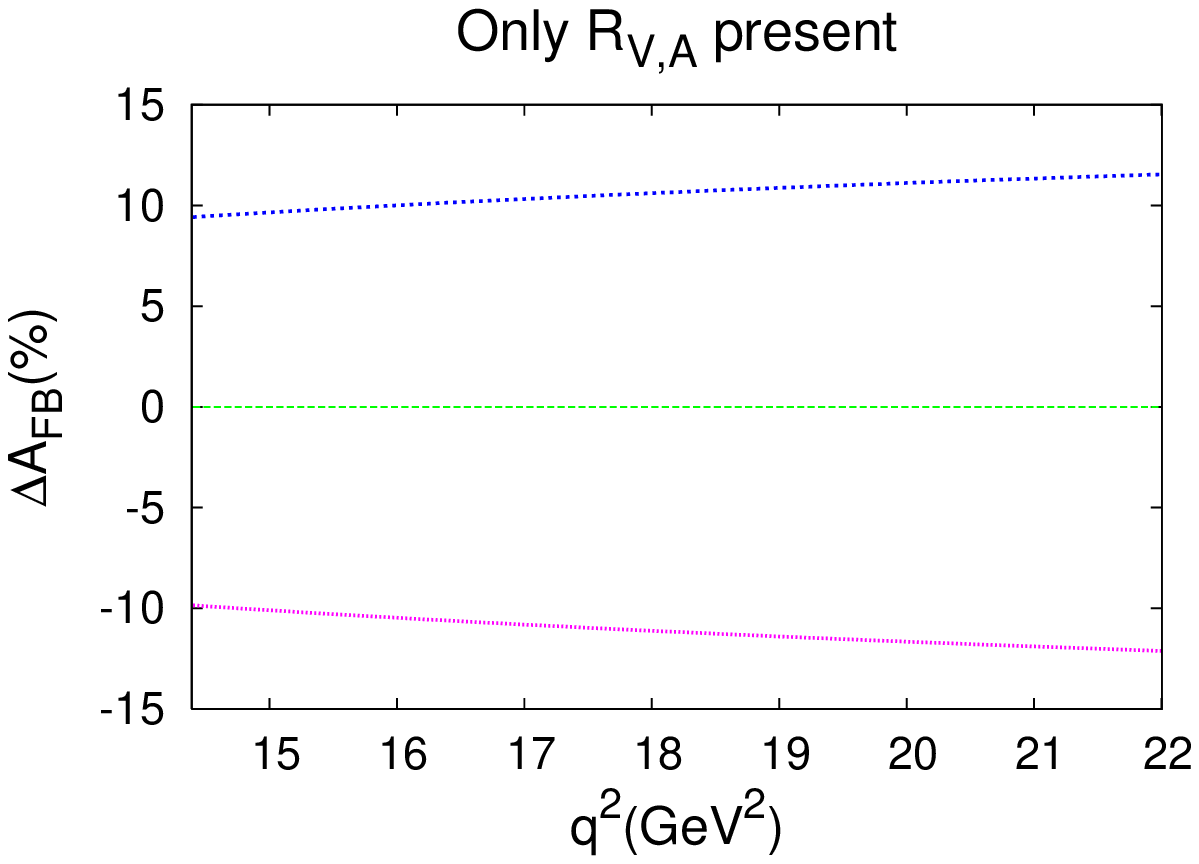}
\caption{The left (right) panels of the figure show $A_{\rm CP}(q^2)$
  and $\Delta A_{FB}$ for $\BXsmumu$ in the low-$q^2$ (high-$q^2$)
  region, in the scenario where only $(R_V, R_A)$ couplings are
  present.  The green line corresponds to the SM prediction. The other
  lines show predictions for some representative values of the NP
  parameters $(R_V,R_A)$. For example, the blue curve in the low-$q^2$
  and high-$q^2$ regions for the $A_{\rm CP}$ plot corresponds to
  $(5.68 e^{i 2.13}, 2.64 e^{-i 0.04})$ and $(4.29 e^{i 1.68}, 4.15
  e^{-i 0.26 })$, respectively, whereas the blue curve in the
  low-$q^2$ and high-$q^2$ regions for the $\Delta A_{\rm FB}$ plot
  corresponds to $(1.80 e^{i 2.91}, 5.45 e^{i 0.90})$ and $(1.69 e^{-i
    3.08}, 6.83 e^{-i 0.91})$, respectively.}
\label{fig:BXsmumu}
}

Fig.~\ref{fig:BXsmumu} shows $A_{\rm CP}(q^2)$ and 
$\Delta A_{FB}(q^2)$ for $\BXsmumu$ in the presence of 
new VA couplings. We make the following observations:

\begin{itemize}

\item When only $R_{V,A}$ couplings are present, $A_{\rm CP}(q^2)$ can
  be enhanced up to 6\% at low $q^2$.  On the other hand, its value at
  high $q^2$ can be as high as 12\%.  $A_{\rm CP}(q^2)$ can have
  either sign at both low and high $q^2$.  At high $q^2$, the
  magnitude of $A_{\rm CP}(q^2)$ is almost independent of $q^2$.

\item When only $R'_{V,A}$ couplings are present, $A_{\rm CP}(q^2)$
  cannot be enhanced above the SM value. This is because $R'_{V,A}$
  couplings do not contribute to the numerator of $A_{\rm CP}(q^2)$ in
  Eq.~(\ref{Acp-Xsmumu}). They can only affect the DBR, which may be
  enhanced by up to 50$\%$, thus decreasing $A_{\rm CP}(q^2)$.

\item In the presence of $R_{V,A}$ couplings, $\Delta A_{FB}$ can be
  enhanced up to 3\% at low $q^2$. At high $q^2$, the enhancement can
  be up to 12\%. The impact of $R'_{V,A}$ couplings is negligible ($<
  1\%$).

\end{itemize}

\section{\boldmath $\Bsmumugamma$
\label{Bsmumugamma}}

Although $\Bsmumugamma$ requires the emission of an additional photon
as compared to $\Bsmumu$, which suppresses the branching ratio (BR) by
a factor of $\alpha_{em}$, the photon emission also frees it from
helicity suppression, making its BR much larger than $\Bsmumu$.  The
SM prediction for the BR in the range $q^2 \le 9.5$ GeV$^2$ and $q^2
\ge 15.9$ GeV$^2$ is $\approx 18.9 \times 10^{-9}$
\cite{Melikhov:2004mk}.  As argued in Ref.~\cite{Alok:2010zd}, if we
choose $2$ GeV$^2 \le q^2 \le 6$ GeV$^2$ and $14.4$ GeV$^2 \le q^2 \le
25$ GeV$^2$ as the low-$q^2$ and high-$q^2$ regions, respectively,
then the dominating contribution comes from the diagrams in which the
final-state photon is emitted either from the $b$ or the $s$ quark,
and the $\Bsmumugamma$ decay is governed by the same $\bsmumu$
effective Hamiltonian as the other decays considered in this paper.

The CP asymmetry in $\Bsmumugamma$ is given in Eq.~(\ref{Acp-Xsmumu}),
where $dB/dx_\gamma$ and $d\overline{B}/dx_\gamma$ are the DBRs of
$\Bsmumugamma$ and its CP-conjugate $B^0_s \to \mu^+ \mu^- \gamma$,
respectively. The expression for $(dB/dx_\gamma)$ has been given in
Ref.~\cite{Alok:2010zd}. The CP asymmetry in $A_{FB}$ is given in
Eq.~(\ref{AFB-Xsmumu}), where the definition of $A_{FB}$ is given in
Ref.~\cite{Alok:2010zd}, and $\overline{A}_{FB}$ is the analogous
quantity for the CP-conjugate decay.

The CP asymmetry in the DBR of $B_s \to \mu \mu \gamma$ was studied in
Refs.~\cite{Balakireva:2009kn,Balakireva:2010zz}, albeit only for the
new-physics cases where $C_7 = -C_7^{\rm SM}, C_9 = -C_9^{\rm SM}$ and
$C_{10} = -C_{10}^{\rm SM}$, and naturally only for VA operators.
Here, we include a complete discussion of the possible enhancement of
the asymmetry for all allowed values of $C_9$ and $C_{10}$, and in the
presence of SP and T operators.  In addition, we study the
CP-violating asymmetry in $A_{FB}$, which also turns out to give
possibly significant NP signals.

\FIGURE[t]{
\includegraphics[width=0.4\linewidth]{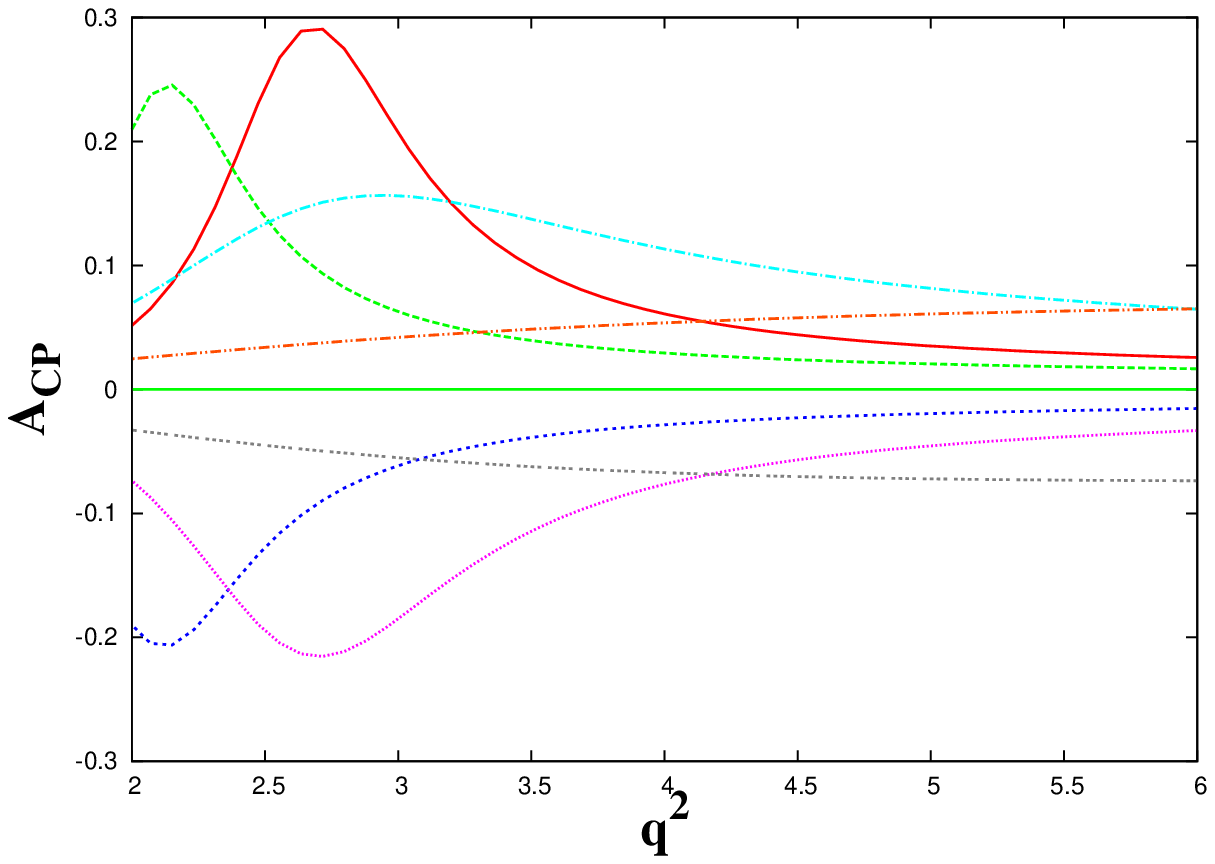} 
\includegraphics[width=0.4\linewidth]{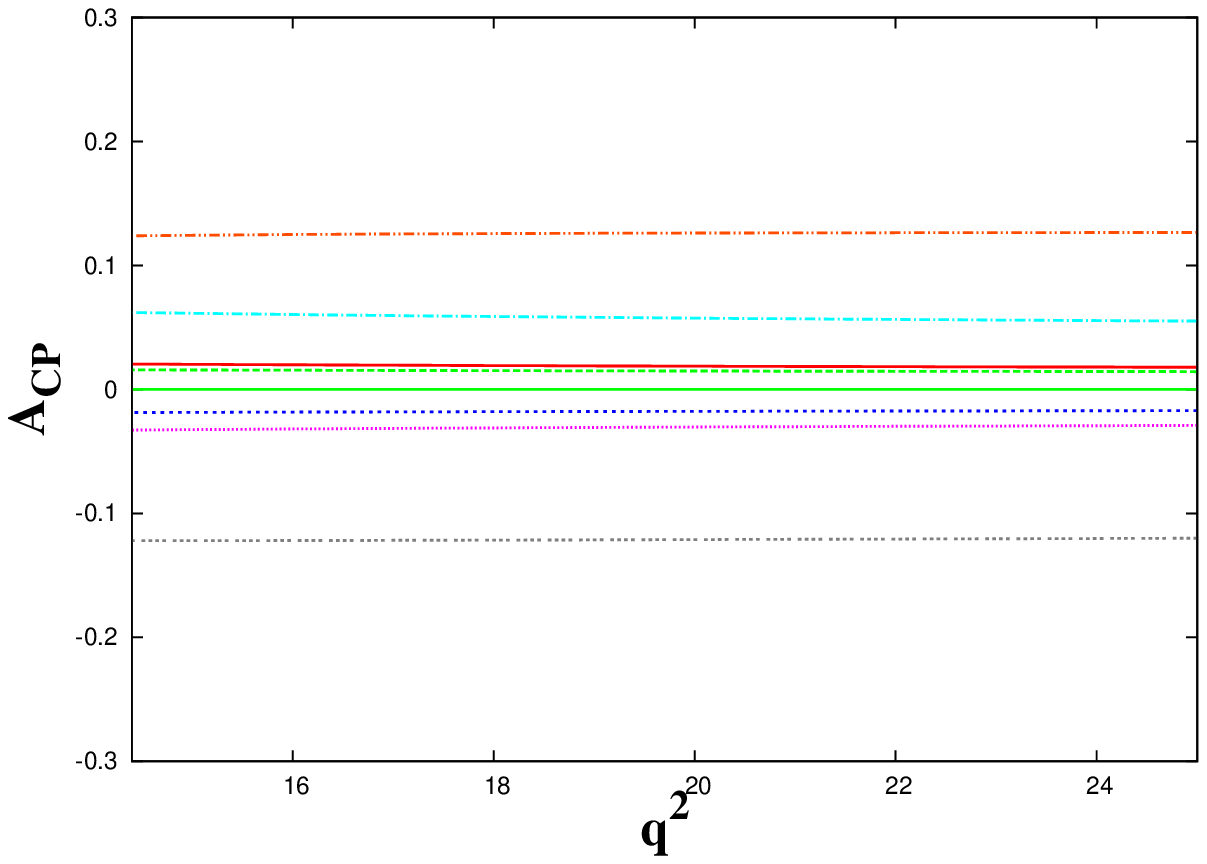}
\includegraphics[width=0.4\linewidth]{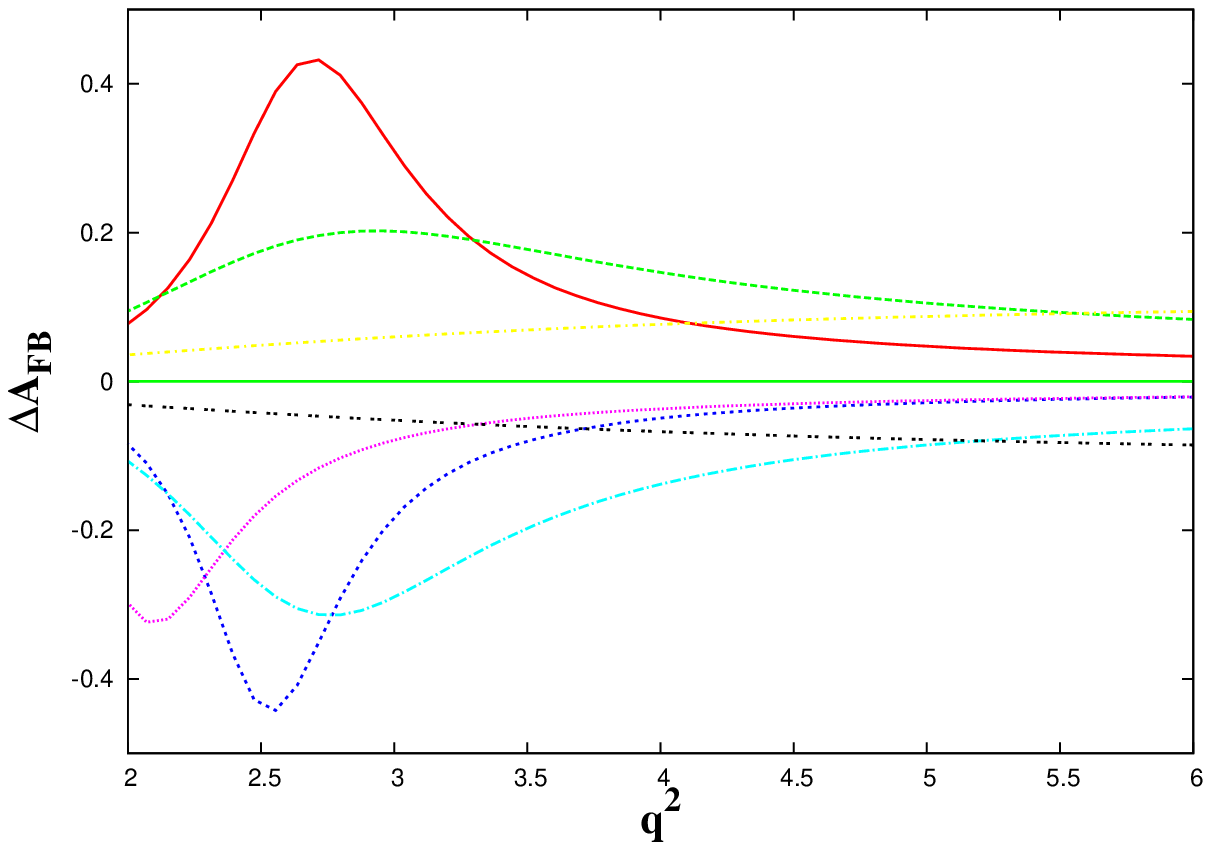} 
\includegraphics[width=0.4\linewidth]{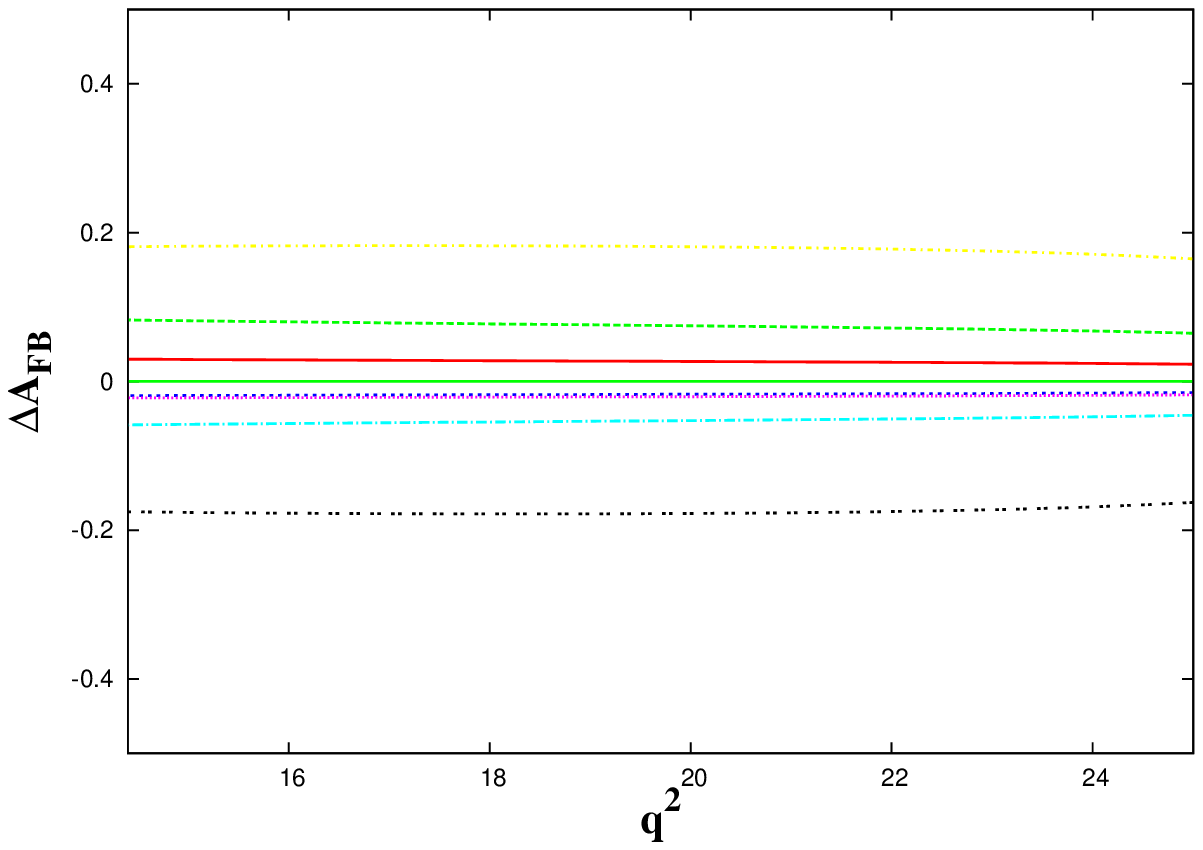}
\caption{The left (right) panels of the figure show $A_{\rm CP}(q^2)$
  and $\Delta A_{FB}$ for $\Bsmumugamma$ in the low-$q^2$ (high-$q^2$)
  region, in the scenario where only $(R_V, R_A)$ couplings are
  present.  For example, the blue curve in the low-$q^2$ and
  high-$q^2$ regions for the $A_{\rm CP}$ plot corresponds to $(2.95
  e^{-i 0.38}, 4.56 e^{-i 0.04})$, whereas the blue curve in the
  low-$q^2$ and high-$q^2$ regions for the $\Delta A_{\rm FB}$ plot
  corresponds to $(1.60 e^{-i 0.08}, 4.14 e^{-i 0.12})$.  }
\label{fig:bsmumugamma}
}

Fig.~\ref{fig:bsmumugamma} shows $A_{\rm CP}(q^2)$ and $\Delta
A_{FB}(q^2)$ for $\Bsmumugamma$ in the presence of new VA
couplings. We make the following observations:

\begin{itemize}

\item When only $R_{V,A}$ couplings are present, at low $q^2$ the
  magnitude of $A_{\rm CP}(q^2)$ can be enhanced up to 30\% at certain
  $q^2$ values.  At high $q^2$, the magnitude of $A_{\rm CP}(q^2)$ is
  almost independent of $q^2$, and can be enhanced to about 13\%.  The
  asymmetry can have either sign at both low and high $q^2$.

\item When only $R'_{V,A}$ couplings are present, $A_{\rm CP}(q^2)$
  cannot be enhanced in magnitude to more than 1.5\% at low $q^2$, or
  more than 3\% at high $q^2$. The detection of NP of this kind is
  therefore expected to be very difficult in this channel.  When both
  primed and unprimed VA couplings are present, the results are the
  same as those obtained with only $R_{V,A}$ couplings.

\item The behaviour of $\Delta A_{FB}(q^2)$ is similar to that of
  $A_{CP}(q^2)$. This quantity can be enhanced up to 40\% for some
  values in the low-$q^2$ region.  It can be as high as 18\%
  throughout the high-$q^2$ region.  The impact of $R'_{V,A}$
  couplings is negligible ($< 1\%$).

\end{itemize}

The new VA operators can therefore enhance the asymmetries $A_{\rm
  CP}(q^2)$ and $\Delta A_{FB}(q^2)$ in $\Bsmumugamma$ to $\sim 10\%$
throughout the $q^2$ region.  For a branching ratio of $O( 2 \times
10^{-8})$, a measurement of a CP asymmetry of $10\%$ at the $3\sigma$
level would require $\sim 10^{10}$ $B$ mesons. It should therefore be
possible to measure a CP asymmetry at the level of a few per cent at
future colliders such as the Super-$B$ factories
\cite{Browder:2008em,Bona:2007qt,O'Leary:2010af}.

\section{\boldmath $\BKmumu$
\label{BKmumu}}

The CP asymmetry in $\BKmumu$ is defined in a manner similar to that
in Eq.~(\ref{Acp-Xsmumu}), where $dB/dz$ and $d\overline{B}/dz$ are
the DBRs of $\BKmumu$ and its CP-conjugate $B^0_d \to K \mu^+ \mu^-$,
respectively.  The expression for $(dB/dz)$ has been given in
Ref.~\cite{Alok:2010zd}.  A model-independent analysis of the CP
asymmetry in the DBR, with specific chosen values of VA operators, was
carried out in Ref.~\cite{Aliev:2005pwa}.  However, the constraints on
the NP operators, coming from the measured branching ratio of
$\BXsmumu$, were not taken into account.  Here, in addition to taking
these constraints into account, we also consider new SP and T
operators, and extend the analysis to study the CP asymmetry in
$A_{FB}$.
 
The CP asymmetry in $A_{FB}$ is given in Eq.~(\ref{AFB-Xsmumu}), where
the definition of $A_{FB}$ is as given in Ref.~\cite{Alok:2010zd},
while $\overline{A}_{FB}$ is the analogous quantity for the
CP-conjugate decay. Now, the decay mode $\BKmumu$ is unique because
the forward-backward asymmetry of muons is predicted to vanish exactly
in the SM.  This is due to the fact that the $\bdbar \to \bar{K}$
hadronic matrix element does not have any axial-vector contribution.
$\AFB$ can therefore have a nonzero value only if it receives a
contribution from new physics. However, even in the presence of NP,
the expressions in Ref.~\cite{Alok:2010zd} indicate that the only term
contributing to $\Delta A_{FB}(q^2)$ is that with VA+SP NP operators,
and this is suppressed by the factor $m_\mu / m_b$. As a result, one
does not expect a significant enhancement in $\Delta A_{FB}$ from any
Lorentz structure of NP.

\FIGURE[t]{
\includegraphics[width=0.4\linewidth]{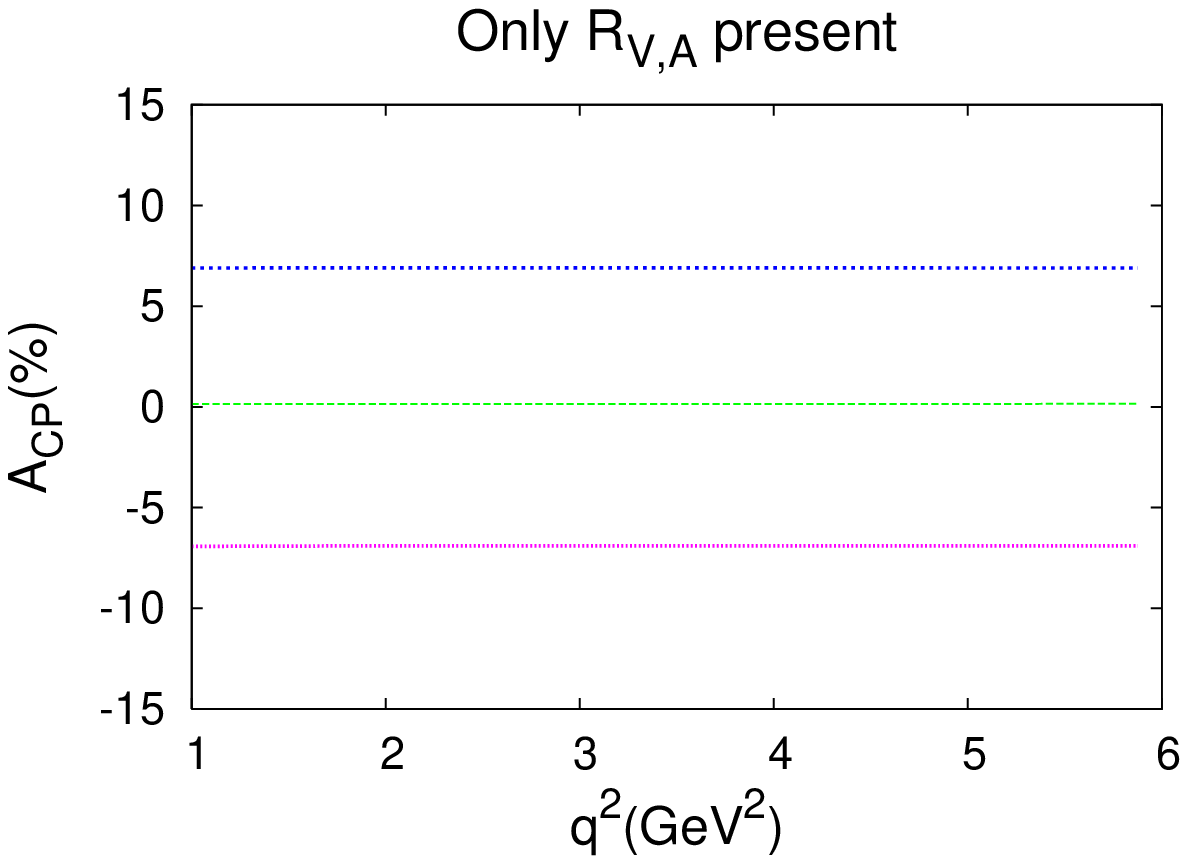} 
\includegraphics[width=0.4\linewidth]{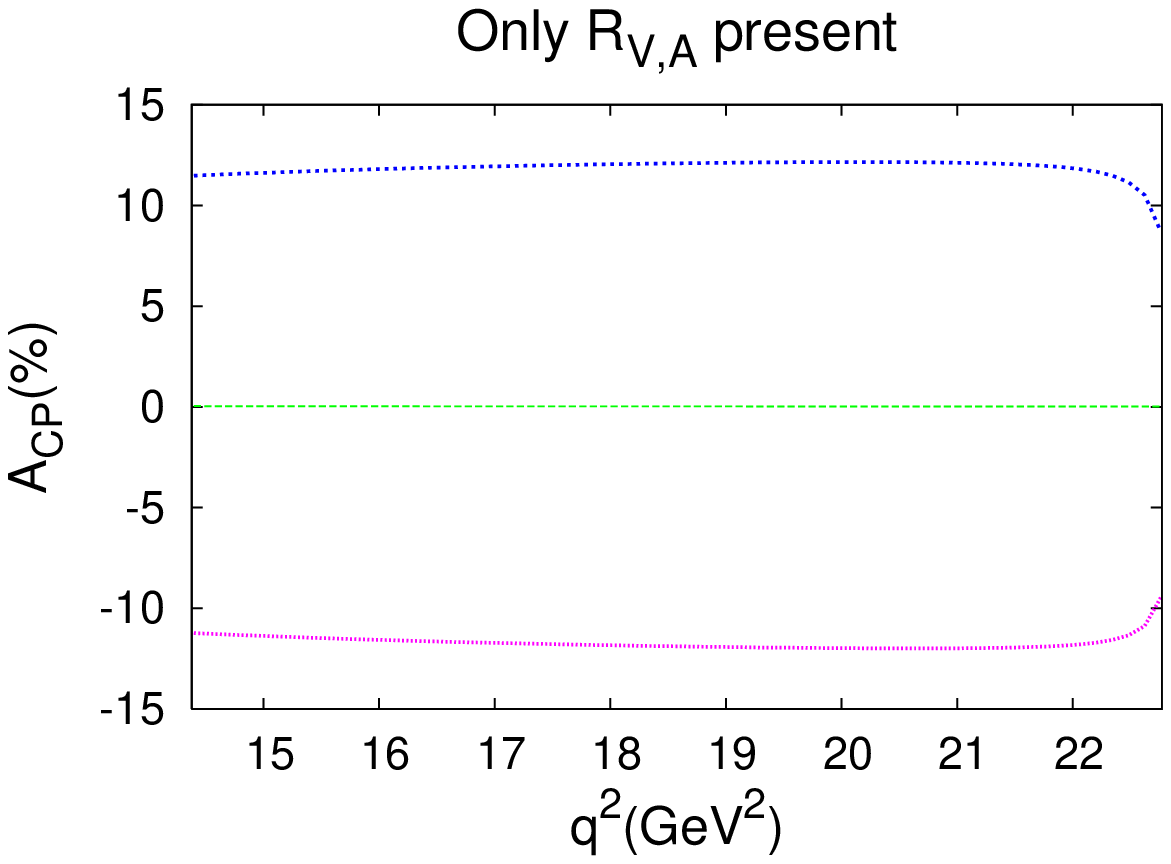}
\caption{The left (right) panel of the figure shows $A_{\rm CP}(q^2)$
  for $\BKmumu$ in the low-$q^2$ (high-$q^2$) region, in the scenario
  where only $(R_V, R_A)$ terms are present.  The green line
  corresponds to the SM prediction. The other lines show predictions
  for some representative values of the NP parameters $(R_V,R_A)$.
  For example, the blue curve in the low-$q^2$ and high-$q^2$ regions
  corresponds to $(5.97 e^{i 2.23}, 3.08 e^{-i 0.05})$ and $(6.47 e^{i
    2.30}, 3.11 e^{i 0.48 })$, respectively.
\label{fig:acp-bkll-va}}
}

Fig.~\ref{fig:acp-bkll-va} shows $A_{\rm CP}(q^2)$ for $\BKmumu$ in
the presence of new VA couplings.  We make the following observations:
\begin{itemize}

\item When only $R_{V,A}$ couplings are present, $A_{\rm CP}(q^2)$ can
  be enhanced up to 7\% at low $q^2$.  On the other hand, its value at
  high $q^2$ can be as high as 12\%.  $A_{\rm CP}(q^2)$ can have
  either sign at both low and high $q^2$, and its magnitude is almost
  independent of $q^2$ in these regions.

\item When only $R'_{V,A}$ couplings are present, $A_{\rm CP}(q^2)$
  can be enhanced up to 4\% at low $q^2$.  On the other hand, its
  value at high $q^2$ can be as high as 12\%.  $A_{\rm CP}(q^2)$ can
  have either sign at both low and high $q^2$, and its magnitude is
  almost independent of $q^2$ in these regions.

\item When both primed and unprimed VA couplings are present, the
  results are the same as those obtained with only $R_{V,A}$
  couplings.


\end{itemize}

For a $\BKmumu$ branching ratio of $O(0.5 \times 10^{-6})$, 
a measurement of a CP asymmetry of $1\%$ at the $3\sigma$ level would 
require $\sim 10^{11}$
B mesons.  It should therefore be possible to measure a CP asymmetry
at the level of a few per cent at future colliders such as the
Super-$B$ factories \cite{Browder:2008em,Bona:2007qt,O'Leary:2010af}.

\section{\boldmath $\BKstarmumu$ 
\label{BKstarmumu}}

The complete three-angle distribution for the decay $\bar{B}^0\to
\bar{K}^{*0} (\to K^-\pi^+)\mu^+\mu^-$ in the presence of NP can be
expressed in terms of $q^2$, two polar angles $\theta_\mu$ and
$\theta_{K}$, and the azimuthal angle $\phi$ between the planes of the
dimuon and $K \pi$ decays:
\bea
\label{ADKst}
&& \frac{d^4\Gamma^{\bar{B}}}{dq^2d\cos{\theta_\mu} d\cos{\theta_{K}} d\phi } 
= N_F
\times \nl
\Bigg\lbrace &&  \cos^2{\theta_{K}} \Big(I^0_1 + I^0_2 \cos{2 \theta_\mu} + I^0_3 \cos{ \theta_\mu} \Big) + \sin^2{\theta_{K}} \Big(I^T_1   + I^T_2  \cos{2 \theta_\mu}   + I^T_3  \cos{ \theta_\mu} \nl &&+ I^T_4  \sin^2{\theta_\mu} \cos{2 \phi}+ I^T_5  \sin^2{\theta_\mu} \sin{2 \phi} \Big) + \sin{2\theta_{K}}\Big( I^{LT}_1  \sin{2\theta_\mu}\cos{ \phi} \nl &&  + I^{LT}_2  \sin{2\theta_\mu}\sin{ \phi}  + I^{LT}_3  \sin{\theta_\mu}\cos{ \phi} + I^{LT}_4  \sin{\theta_\mu}\sin{ \phi}\Big)  \Bigg\rbrace \; .
\eea
The expressions for the normalization $N_F$ and the $I$'s are given in
Ref.~\cite{Alok:2010zd}.  The $I$'s are functions of the couplings,
kinematic variables and form factors.  The definitions of the angles
in $\BKstarmumu$ involve the directions of the $\mu^+$ and
$\overline{K}^*$.  For the CP-conjugate decay $B^0 \to K^{*0} (\to K^+
\pi^-)\mu^+\mu^-$, one defines these angles relative to the directions
of the $\mu^-$ and $K^*$.  The $\bar{I}$'s can be obtained from the
$I$'s by replacing $\theta_\mu \to \theta_\mu - \pi$ and $\phi \to
-\phi$, and changing the signs of the weak phases.

The CP asymmetries in the branching ratio and forward-backward
asymmetry were analyzed in Ref.~\cite{Kruger:2000zg} with the
measurement of $B \to X_s \gamma$ and the limit on the $\BKstarmumu$
branching ratio available then.  An analysis of CP asymmetries in
$\BKstarmumu$ in the low-$q^2$ region was also performed earlier in
Ref.~\cite{Altmannshofer:2008dz}.  We extend this analysis by
including T operators, and present our results for all asymmetries, in
both the low-$q^2$ and high-$q^2$ regions.

A detailed discussion of the CP-conserving observables in this decay
distribution can be found in Ref.~\cite{Alok:2010zd}.  In this section
we consider the direct CP asymmetries in the differential branching
ratio (DBR), the forward-backward asymmetry $A_{FB}$, the longitudinal
polarization fraction $f_L$, and the angular asymmetries $A_T^{(2)}$
and $A_{LT}$. We also examine the triple-product CP asymmetries
$A_T^{(im)}$ and $A^{(im)}_{LT}$, which were not considered
in Ref.~\cite{Alok:2010zd} since they identically vanish in the
CP-conserving limit (no strong or weak phases), 
regardless of the presence of NP.

\subsection{\boldmath Direct CP asymmetries in the DBR and $A_{FB}$}

The direct CP asymmetry in the differential branching ratio is defined as
\bea
\label{ACPBt-exp}
A_{CP}(q^2) &=& \frac{(d\Gamma^{\bar{B}}/dq^2)- (d\Gamma^{B}/dq^2)}
{(d\Gamma^{\bar{B}}/dq^2) + (d\Gamma^{B}/dq^2)} \; ,
\eea
where 
\bea
\frac{d\Gamma^{\bar{B}}}{dq^2 } &=& \frac{8 \pi N_F}{3} 
(A^{\bar{B}}_{L}+A^{\bar{B}}_{T}) \; .
\eea
Here the longitudinal and transverse polarization amplitudes
$A^{\bar{B}}_L$ and $A^{\bar{B}}_T$  are obtained from Eq.~(\ref{ADKst}):
\bea
\label{HL}
A^{\bar{B}}_L &=& \Big(I^0_1  - \frac{1}{3} I^0_2 \Big),\quad A^{\bar{B}}_T = 2 \Big(I^T_1  - \frac{1}{3} I^T_2 \Big) ~.
\eea
The expressions for $A^{B}_L$ and $A^{B}_T$ of the CP-conjugate mode
can be obtained by replacing the $I$'s with $\bar{I}$'s.

The forward-backward asymmetry in $\BKstarmumu$ has recently been
measured, and shows features that may indicate a deviation from the
SM. This measured quantity is actually the CP-averaged
forward-backward asymmetry $A_{FB}$.  However, the difference between
the measurement of this quantity in $\BKstarmumu$ and its CP-conjugate
mode may also reveal the presence of NP.  This CP asymmetry is
quantified as
\bea
\label{FBAnew}
\Delta A_{FB}(q^2) &=&  A^{\bar{B}}_{FB}(q^2) + A^{B}_{FB}(q^2)  \; ,
\eea
where 
\bea
\label{FBA}
A^{\bar{B}(B)}_{FB}(q^2) &=&\frac{\int^1_0 d\cos{\theta_\mu}
\frac{d^2\Gamma^{\bar{B}(B)}}{dq^2d\cos{\theta_\mu}  }
-\int^0_{-1} d\cos{\theta_\mu}
\frac{d^2\Gamma^{\bar{B}(B)}}{dq^2d\cos{\theta_\mu}  }}
{\int^1_0 d\cos{\theta_\mu} \frac{d^2\Gamma^{\bar{B}(B)}}
{dq^2d\cos{\theta_\mu}  }+\int^0_{-1} d\cos{\theta_\mu}
\frac{d^2\Gamma^{\bar{B}(B)}}{dq^2d\cos{\theta_\mu}  }} \; .
\eea
It can be obtained by integrating over the two angles $\theta_{K}$ and
$\phi$ in Eq.~(\ref{ADKst}). 

Fig.~\ref{fig:ACPBr-vavapKstmumu} shows $A_{CP}(q^2)$ and $\Delta
A_{FB}(q^2)$ for $\BKstarmumu$ in the presence of new VA couplings.
We make the following observations:
\begin{itemize}

\item If only $R_{V,A}$ couplings are present, $A_{CP}(q^2)$ can be
  enhanced up to 5\% at low $q^2$, and up to 14 \% at high
  $q^2$. $\Delta A_{FB}(q^2)$ can be enhanced up to 3\% at low $q^2$,
  and up to 11 \% at high $q^2$. Both $A_{CP}(q^2)$, and $\Delta
  A_{FB}(q^2)$ can have either sign at both low and high $q^2$.

 \item If only $R'_{V,A}$ couplings are present, $A_{CP}(q^2)$ can be
   enhanced up to 3\% at low $q^2$, and up to 7\% at high $q^2$.
   $\Delta A_{FB}(q^2)$ can be enhanced up to 1\% at low $q^2$, and up
   to 4 \% at high $q^2$. Both $A_{CP}(q^2)$, and $\Delta A_{FB}(q^2)$
   can have either sign at both low and high $q^2$.

\item When both primed and unprimed VA couplings are present,
  $A_{CP}(q^2)$ can be enhanced up to 9\% at low $q^2$, and up to 14
  \% at high $q^2$. $\Delta A_{FB}(q^2)$ can be enhanced up to 6\% at
  low $q^2$, and up to 19 \% at high $q^2$. Both $A_{CP}(q^2)$, and
  $\Delta A_{FB}(q^2)$ can have either sign at both low and high $q^2$
  (see Fig.~\ref{fig:ACPBr-vavapKstmumu}).

\end{itemize}

These observations are consistent with the rough expectations in
Ref.~\cite{Kruger:2000zg} about the effect of VA operators.

\FIGURE[t]{
\includegraphics[width=0.4\linewidth]{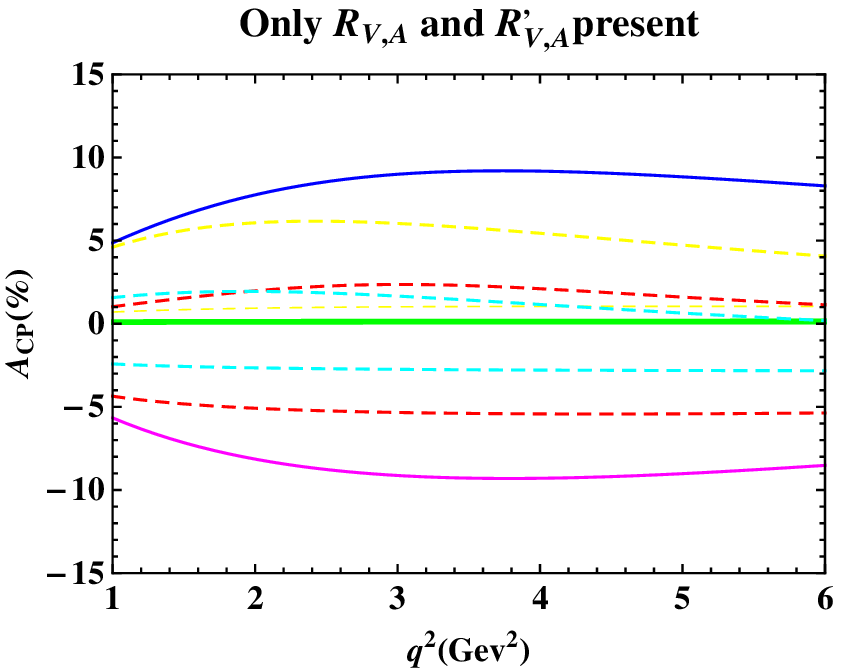}
\includegraphics[width=0.4\linewidth]{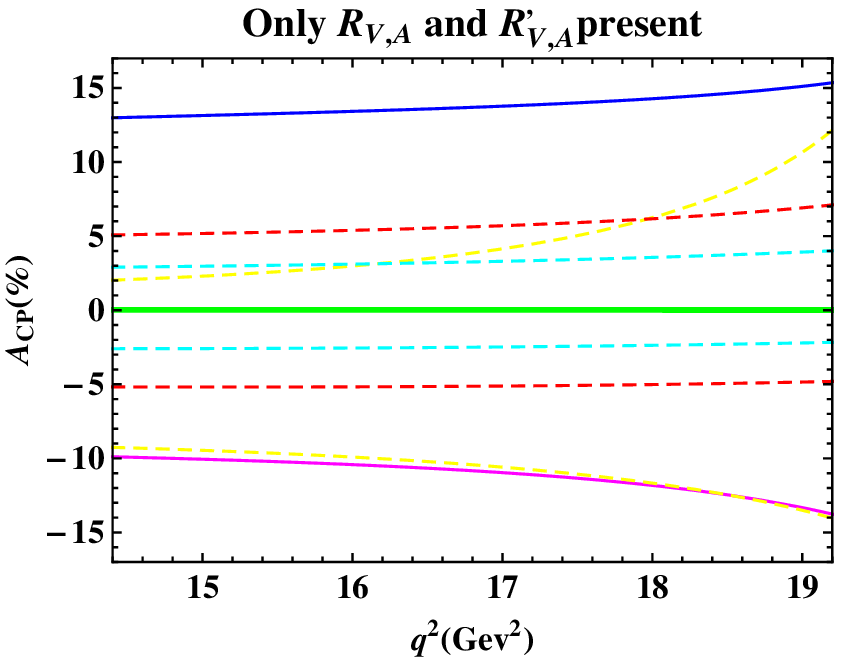} \\
\includegraphics[width=0.4\linewidth]{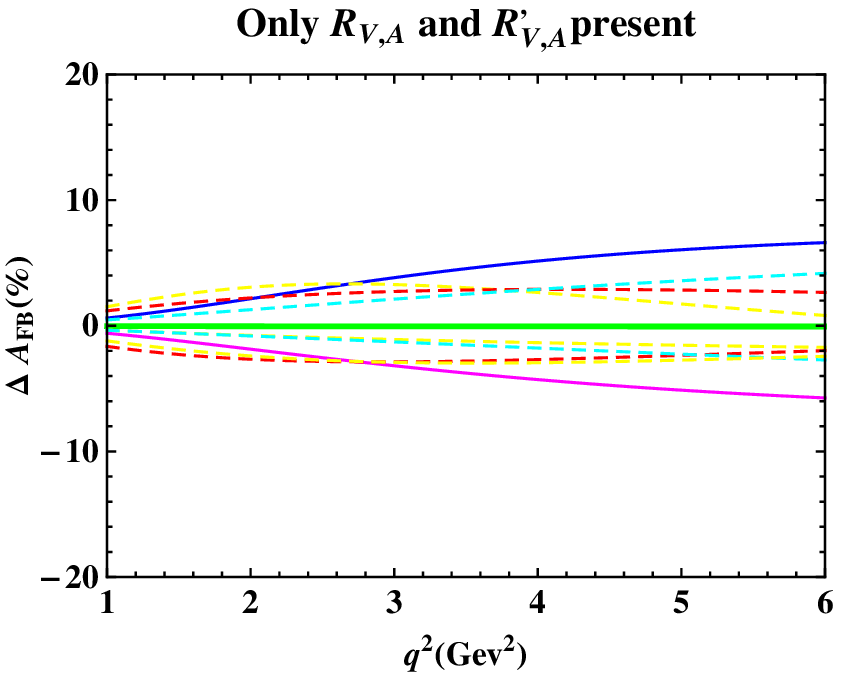}
\includegraphics[width=0.4\linewidth]{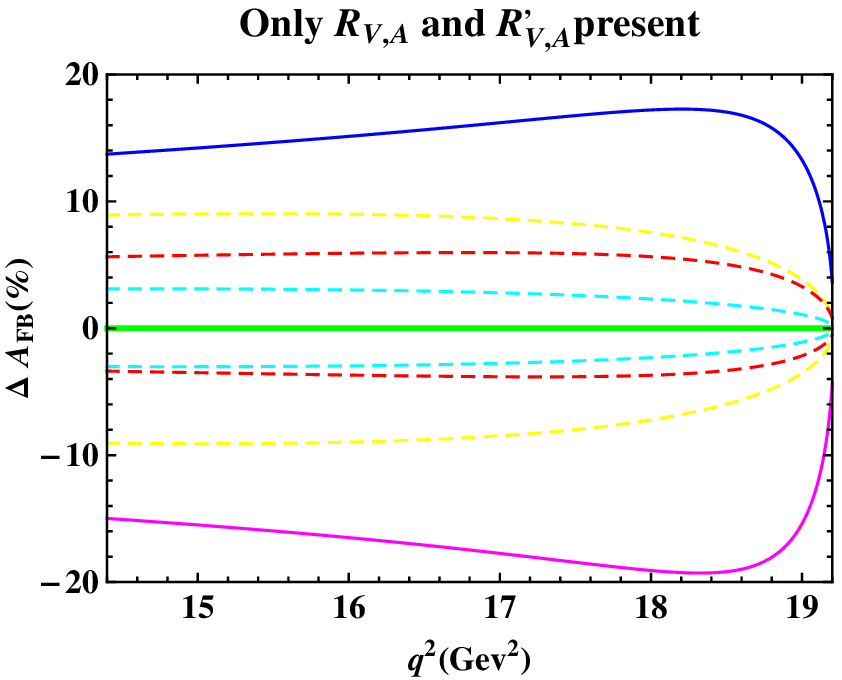}
\caption{The left (right) panels of the figure show $A_{CP}(q^2)$ and
  $\Delta A_{FB}(q^2)$ for $\BKstarmumu$ in the low-$q^2$ (high-$q^2$)
  region, in the scenario where $(R_V, R_A, R^\prime_V, R^\prime_A)$
  terms are all present. The green line corresponds to the SM
  prediction. The other lines show predictions for some representative
  values of the NP parameters.  For example, the blue curve for
  $A_{CP}(q^2)$ in the low-$q^2$ and high-$q^2$ regions corresponds to
  $ (2.77 e^{i 1.83}, 2.08 e^{i 0.5}, 3.8 e^{i 0.08}, 1.23 e^{-i
    2.74})$ and $(5.88 e^{i 2.29}, 1.66 e^{i 0.82}, 3.49 e^{i 0.36},
  1.02 e^{i 0.98})$, respectively. The blue curve for $\Delta
  A_{FB}(q^2)$ in the low-$q^2$ and high-$q^2$ regions corresponds to
  $ (1.56 e^{-i 2.59}, 1.80 e^{-i 0.35}, 4.23 e^{i 0.67}, 1.29 e^{i
    1.43})$ and $(3.21 e^{i 2.61}, 1.38 e^{i2.26}, 5.55 e^{i 0.69},
  3.03 e^{i 1.92})$, respectively.
  \label{fig:ACPBr-vavapKstmumu}}
}

\subsection{\boldmath Direct CP asymmetry in the polarization fraction $f_L$}

The CP asymmetry in the longitudinal polarization fraction $f_L$ is defined as
\bea
\label{ACPpol-exp}
\Delta f_L = f^{\bar{B}}_L-f^{B}_L \; ,
\eea
where
\bea
\label{flft}
f^{\bar{B}(B)}_L &=& \frac{A^{\bar{B}(B)}_L}{A^{\bar{B}(B)}_L+A^{\bar{B}(B)}_T} ~.
\eea

Fig.~\ref{fig:ACPfl-va} shows $\Delta f_L(q^2)$ for $\BKstarmumu$ in
the presence of new VA couplings.  We make the following observations:
\begin{itemize}

\item If only $R_{V,A}$ couplings are present, $\Delta f_L(q^2)$ can
  be enhanced up to 2\% at very low $q^2$. On the other hand, $\Delta
  f_L(q^2)$ is almost the same as the SM at high $q^2$. It can have either
  sign at both low and high $q^2$.

\item If only $R'_{V,A}$ couplings are present, $\Delta f_L(q^2)$ can
  be enhanced up to 2\% at both low and high $q^2$. It can have either
  sign at both low and high $q^2$.

\item When both primed and unprimed VA couplings are present, $\Delta
  f_L(q^2)$ can be enhanced up to 9\% at low $q^2$, and up to 6\% at
  high $q^2$. It can have either sign at both low and high $q^2$ (see
  Fig.~\ref{fig:ACPfl-va}).

\end{itemize}

\FIGURE[t]{
\includegraphics[width=0.4\linewidth]{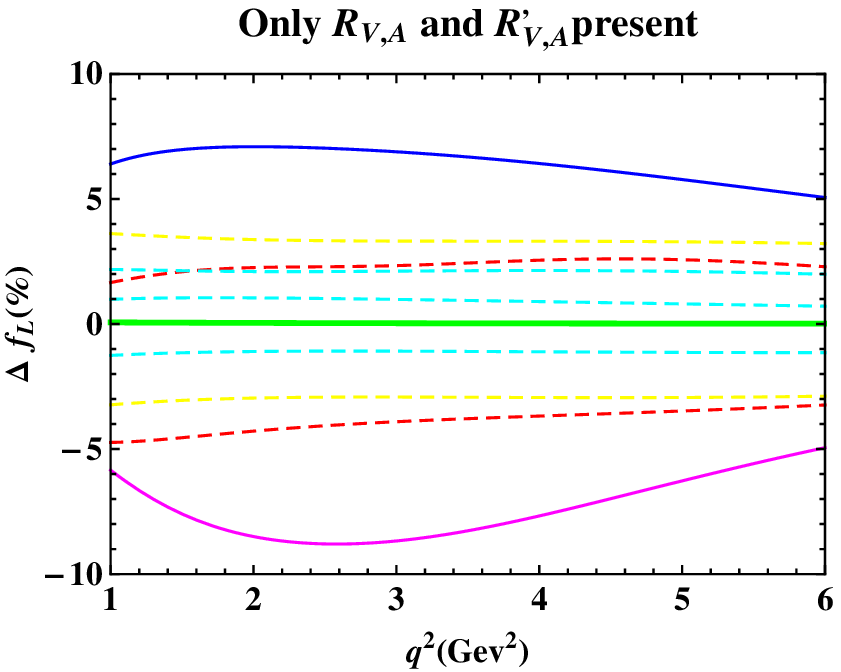}
\includegraphics[width=0.4\linewidth]{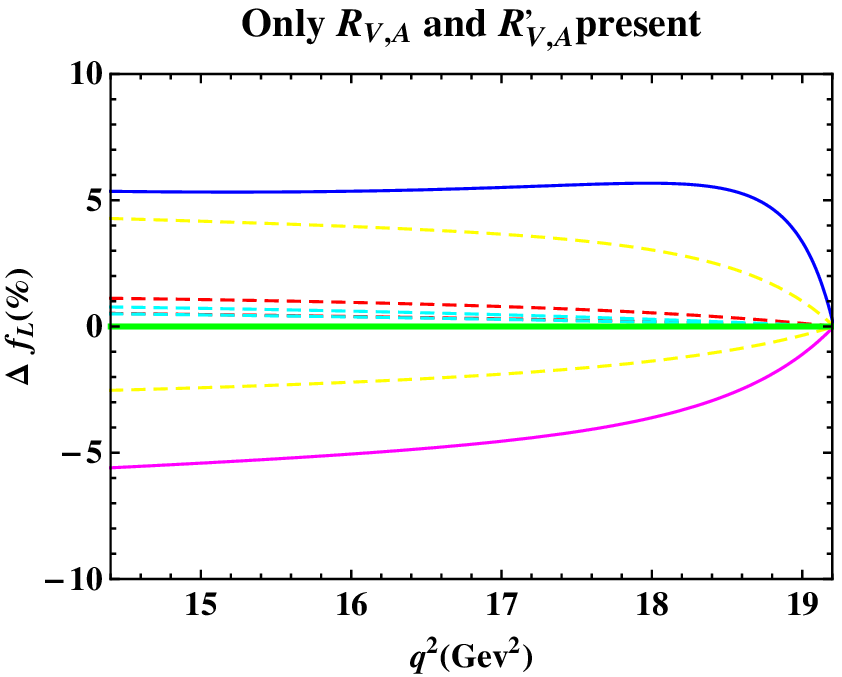}
\caption{The left (right) panel of the figure shows $\Delta f_L(q^2)$
  for $\BKstarmumu$ in the low-$q^2$ (high-$q^2$) region, in the
  scenario where $(R_V, R_A, R^\prime_V, R^\prime_A)$ terms are all
  present.  For example, the blue curve in the low-$q^2$ and
  high-$q^2$ regions corresponds to $ (2.78 e^{i 2.98}, 2.19 e^{-i
    0.77},6.91 e^{-i 0.29}, 3.34 e^{-i 0.56})$.
\label{fig:ACPfl-va}}
}

\subsection{\boldmath Direct CP asymmetries in the angular asymmetries $A_T^{(2)}$ and $A_{LT}$}

The transverse asymmetry $A_T^{(2) \bar{B} (B)}$ is defined
\cite{kruger-matias} through the double differential decay rate as
\bea
\label{doubDR3}
\frac{d^2\Gamma^{\bar{B}(B)}}{dq^2 d\phi } &=&\frac{1}{2\pi} \frac{d\Gamma^{\bar{B}(B)}}{dq^2  }
\Big[ 1+ f^{\bar{B}(B)}_T \left(A^{(2)\bar{B}(B)}_T \cos{2 \phi} + A^{(im)\bar{B}(B)}_T \sin{2 \phi}\right)
\Big] \; .
\eea
It can be obtained by integrating Eq.~(\ref{ADKst}) over the two polar
angles $\theta_{\mu}$ and $\theta_{K}$. Here $A^{(im)\bar{B}(B)}_T$ is
a triple product, and is discussed separately below.  In terms of the
coupling constants and matrix elements defined in
Ref.~\cite{Alok:2010zd}, $A^{(2)\bar{B}(B)}_T$ can be expressed as
\bea
 \label{AT2}
A^{(2)\bar{B}}_T &=& \frac{4 I^T_4}{3 A^{\bar{B}}_T} ~, \quad  A^{(2)B}_T = \frac{4 \bar{I}^T_4}{3 A^B_T} ~.
\eea

While $A^{(2)\bar{B}}_T$ ( $A^{(2)B}_T$) is finite even in the
CP-conserving limit (and was discussed in Ref.~\cite{Alok:2010zd}), a
CP asymmetry may be defined through the difference
\be
\Delta A_T^{(2)} \equiv A_T^{(2)\bar{B}} - A_T^{(2) B} \; . 
\ee

Fig.~\ref{fig:ACPAT2-va} shows $\Delta A_T^{(2)}$ for $\BKstarmumu$ in
the presence of new VA couplings.  We make the following observations:
\begin{itemize}

\item If only $R_{V,A}$ couplings are present, $\Delta A_T^{(2)}$
  cannot be enhanced more than 1\% at both low and high $q^2$.  It can
  have either sign at both low and high $q^2$.

\item If only $R'_{V,A}$ couplings are present, $\Delta A_T^{(2)}$ can
  be enhanced up to 4\% at low $q^2$, and up to 6\% high $q^2$. It can
  have either sign at both low and high $q^2$.

\item When both primed and unprimed VA couplings are present, $\Delta
  A_T^{(2)}$ can be enhanced up to 11\% at low $q^2$, and up to 12\%
  at high $q^2$. It can have either sign at both low and high $q^2$
  (see Fig.~\ref{fig:ACPAT2-va}).

\end{itemize}

\FIGURE[t]{
\includegraphics[width=0.4\linewidth]{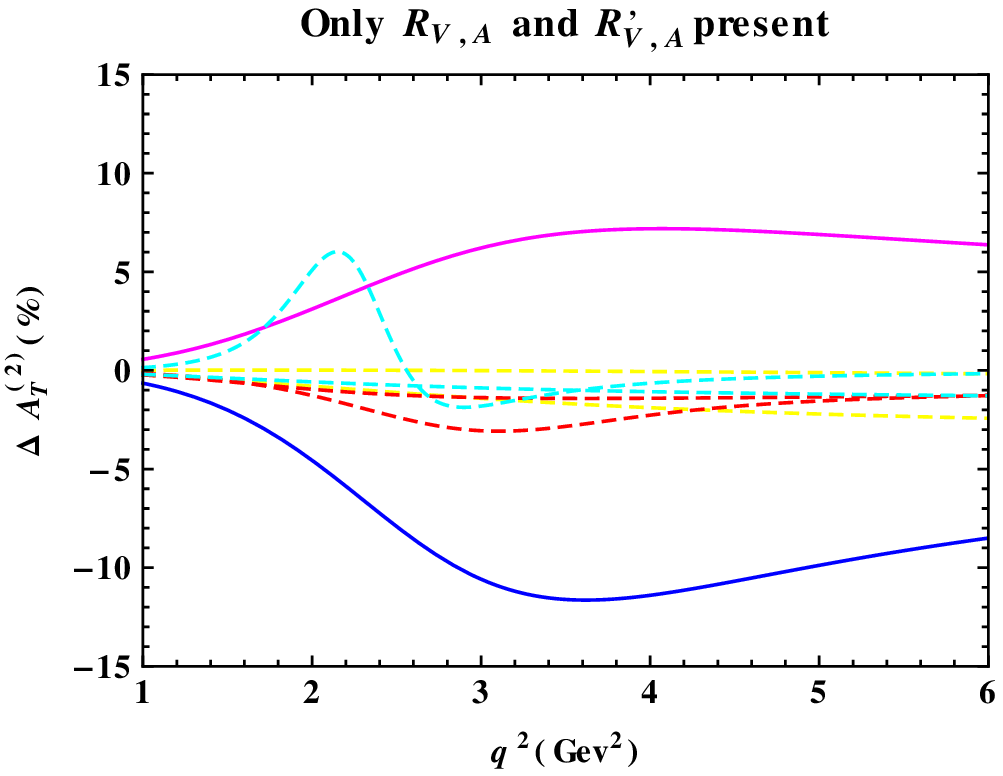}
\includegraphics[width=0.4\linewidth]{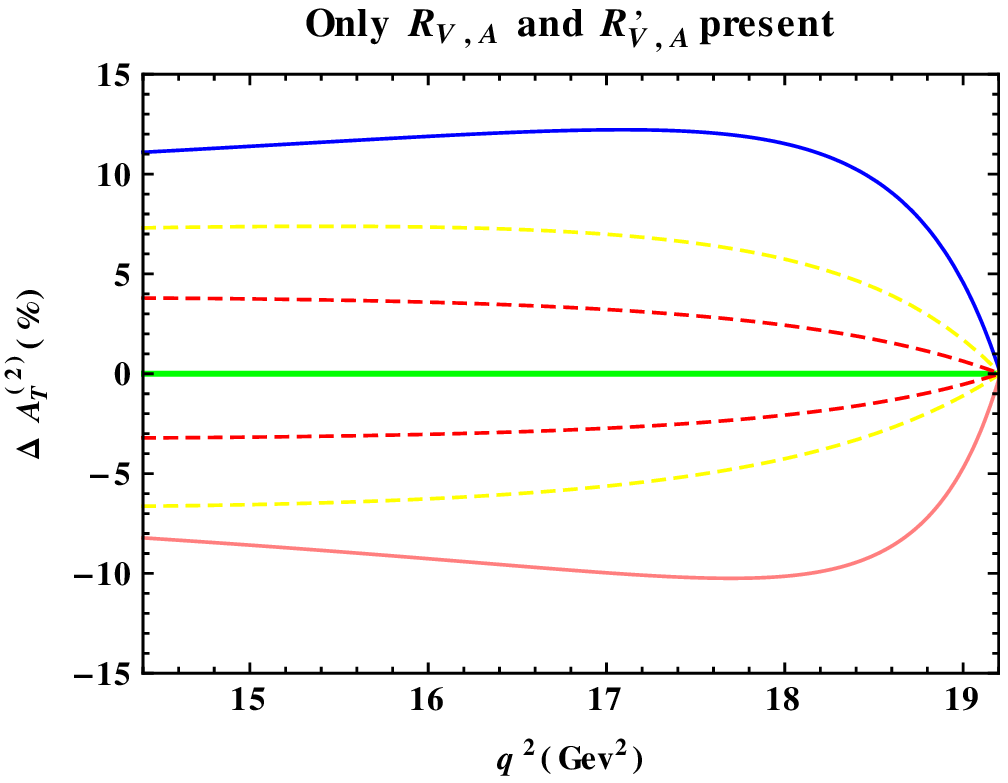}
\caption{The left (right) panel of the figure shows $\Delta
  A_T^{(2)}(q^2)$ for $\BKstarmumu$ in the low-$q^2$ (high-$q^2$)
  region, in the scenario where $(R_V, R_A, R^\prime_V, R^\prime_A)$
  terms are all present. The green line corresponds to the SM
  prediction. The other lines show predictions for some representative
  values of the NP parameters.  For example, the blue curve in the
  low-$q^2$ and high-$q^2$ regions corresponds to $ (0.11 e^{i 2.18},
  2.66 e^{-i 1.31}, 4.3 e^{i 0.03}, 0.23 e^{-i 2.27})$ and $(2.32 e^{i
    2.51}, 4.89 e^{i 1.27}, 3.12 e^{i 0.42}, 0.14 e^{-i 1.55})$,
  respectively.
\label{fig:ACPAT2-va}}
}

The longitudinal-transverse asymmetry $A_{LT}^{\bar{B}(B)}$ is defined
through
\bea
\label{defn-ALT}
\frac{d^2\Gamma_{LT}^{\bar{B}(B)}}{dq^2 d\phi } &=&
\frac{d\Gamma^{\bar{B}(B)}}{dq^2  }
\left(A_{LT}^{(re)\bar{B}(B)} \cos{\phi} + A_{LT}^{(im)\bar{B}(B)} 
\sin{\phi}\right) \; ,
\eea
where
\bea
\frac{d^2\Gamma^{\bar{B}(B)}_{LT}}{dq^2 d\phi } &=& 
\int^1_0  d\cos{\theta_{K}} \frac{d^3\Gamma^{\bar{B}(B)}}{dq^2 d\cos{\theta_{K}} 
d\phi } 
-\int^0_{-1}  d\cos{\theta_{K}} \frac{d^3\Gamma^{\bar{B}(B)}}{dq^2 d\cos{\theta_{K}} d\phi } \; .
\eea
Here $A_{LT}^{(im)\bar{B}(B)}$ is a triple product, and is discussed
separately below. In terms of the coupling constants and matrix
elements defined in Ref.~\cite{Alok:2010zd}, $A^{(re)\bar{B}(B)}_{LT}$ can
be expressed as
\bea
 \label{ALT1-expr1}
  A^{(re)\bar{B}}_{LT} &=& \frac{I^{LT}_3}{4 (A^{\bar{B}}_L + A^{\bar{B}}_T) } ~, \quad 
A^{(re)B}_{LT} =-\frac{\bar{I}^{LT}_3}{4 (A^B_L + A^B_T) } ~.
 \eea
Note that $ A^{(re)B}_{LT}= -A^{(re)\bar{B}}_{LT}$ in the CP-conserving limit. 
Thus, a CP asymmetry may be defined  through the sum
\be
\Delta A_{LT}(q^2) \equiv A_{LT}^{(re) \bar{B}}(q^2) + A_{LT}^{(re) B} (q^2) \; . 
\ee

We now assume the presence of new VA couplings. However, we find that
these couplings cannot enhance $\Delta A_{LT}(q^2)$ to more than 3\%
at both low and high $q^2$.

Note that $\Delta A_{LT}(q^2)$ is related to the observable $A_5^D$ in
Ref.~\cite{Bobeth:2008ij}: $\Delta A_{LT}(q^2) \approx A_5^D/4$.  Our
limit of 3\% on the maximum value of $\Delta A_{LT}(q^2)$ is then
consistent with the limit of 0.07 on the average value $\langle A_5^D
\rangle$ over the low-$q^2$ region, as calculated in
Ref.~\cite{Bobeth:2008ij}.

\subsection{CP-violating triple-product asymmetries}

In this subsection, we consider the triple products (TPs) in the
decays $\bar{B}^0\to \bar{K}^{*0} (\to K^-\pi^+)\mu^+\mu^-$ and $B^0
\to K^{*0} (\to K^+ \pi^-)\mu^+\mu^-$. For the decaying $\bar{B}$
meson, the TP is proportional to $(\hat{n}_K \times \hat{n}_\mu) \cdot
\hat{n}_z$ in its rest frame, where the unit vectors are given in
terms of the momenta of the final-state particles as
\bea
\label{unitvecdef}
\hat{n}_K &=& \frac{\hat{p}_{K^-} \times \hat{p}_{\pi^+} }
{|\hat{p}_{K^-}\times \hat{p}_{\pi^+}|},~~\hat{n}_z = 
\frac{\hat{p}_{K^-}+ \hat{p}_{\pi^+}}{|\hat{p}_{K^-} + \hat{p}_{\pi^+}|},
~~\hat{n}_\mu =\frac{\hat{p}_{\mu^-} \times \hat{p}_{\mu^+} }
{|\hat{p}_{\mu^-} \times \hat{p}_{\mu^+} |} \; .
\eea
In terms of the azimuthal angle $\phi$, one gets 
\bea
\label{unitvec}
\cos{\phi} &=& \hat{n}_K \cdot \hat{n}_\mu\; , \quad
\sin{\phi}= (\hat{n}_K \times \hat{n}_\mu) \cdot \hat{n}_z \; ,
\eea
and hence the quantities that are coefficients of $\sin \phi$ (or of
$\sin 2\phi = 2 \sin\phi \cos \phi$) are the TPs.

As noted above, while the angular distribution for the $\bar{B}$ decay
involves $\phi$, for $B$ it involves $-\phi$. Thus, the CP-violating
triple-product asymmetry is proportional to the {\it sum} of $\bar{B}$
and $B$ TPs.

The first TP is $A^{(im)\bar{B}(B)}_T$, introduced above in
Eq.~(\ref{doubDR3}). In terms of the coupling constants and matrix
elements defined in Ref.~\cite{Alok:2010zd}, $A_T^{(im)\bar{B}(B)}$
can be written as
\bea
 \label{AT2im}
 A^{(im)\bar{B}}_T &=& \frac{4 I^T_5}{3 A^{\bar{B}}_T},\quad A^{(im)B}_T = -\frac{4 \bar{I}^T_5}{3 A^B_T} \; . 
\eea
We observe that $A^{(im)}_T$ depends only on the VA couplings. The
CP-violating triple-product asymmetry is
\bea
\label{ATimAvgdef}
A_T^{(im)} &=& \frac{1}{2} (A_T^{(im)\bar{B}}+ A_T^{(im)B}) ~.
\eea

Fig.~\ref{fig:imAT1-va} shows $A^{(im)}_{T}(q^2)$ for $\BKstarmumu$ in
the presence of new VA couplings.  We make the following observations:
\begin{itemize}

\item If only $R_{V,A}$ couplings are present, $A^{(im)}_{T}(q^2)$ can
  be enhanced up to 5\% at low $q^2$ and can have either sign. On the
  other hand, $A^{(im)}_{T}(q^2)$ is almost same as the SM prediction
  ($\simeq 0$) at high $q^2$.

\item If only $R'_{V,A}$ couplings are present, $A^{(im)}_{T}(q^2)$
  can be enhanced up to 49\% at low $q^2$, and up to 46\% at high
  $q^2$. It can have either sign at both low and high $q^2$.
 
 \item When both primed and unprimed VA couplings are present, the
   results for $A^{(im)}_{T}(q^2)$ are almost the same as those
   obtained with only $R'_{V,A}$ couplings (see
   Fig.~\ref{fig:imAT1-va}).

\end{itemize}

\FIGURE[t]{
\includegraphics[width=0.4\linewidth]{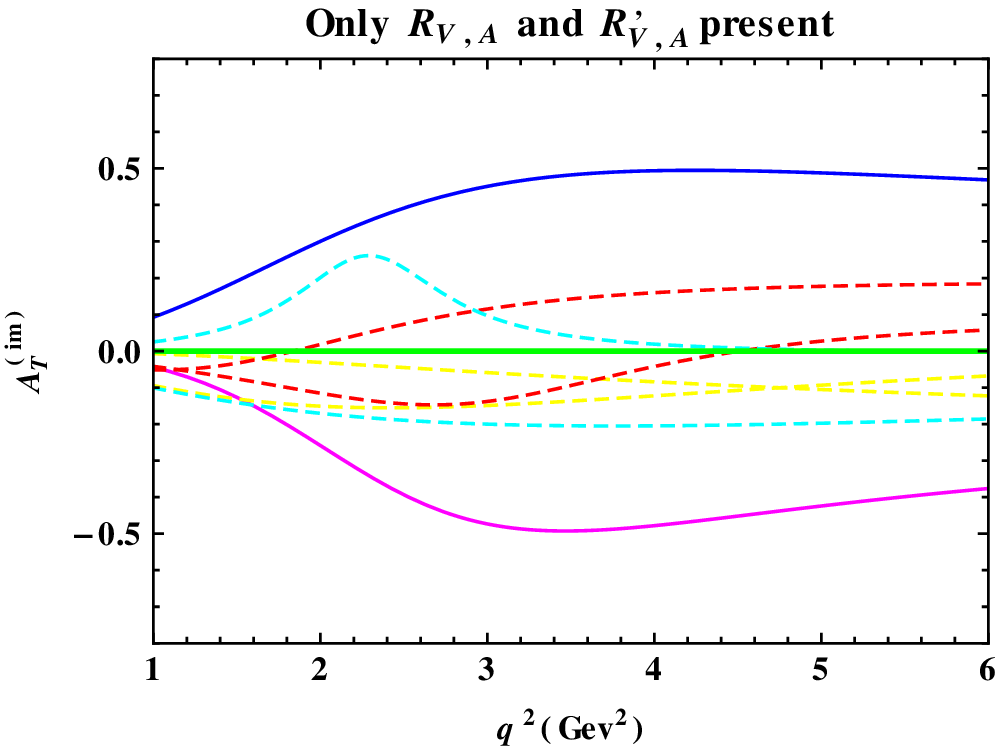}
\includegraphics[width=0.4\linewidth]{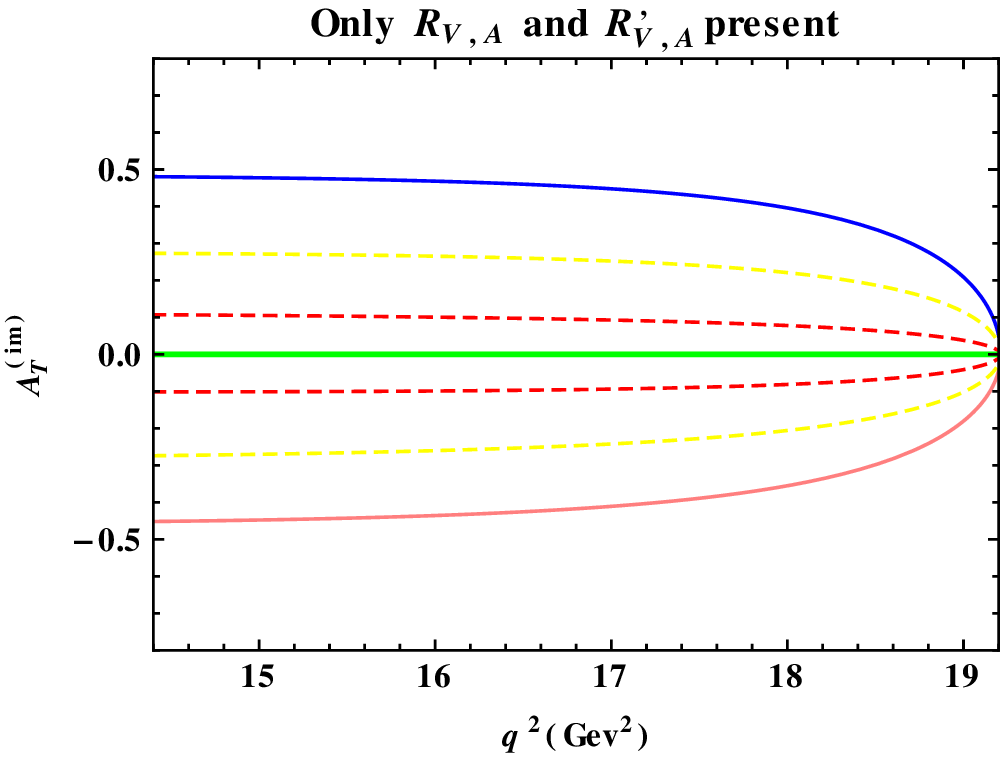}
\caption{The left (right) panel of the figure shows
  $A^{(im)}_{T}(q^2)$ for $\BKstarmumu$ in the low-$q^2$ (high-$q^2$)
  region, in the scenario where $(R_V, R_A, R^\prime_V, R^\prime_A)$
  terms are all present. The green line corresponds to the SM
  prediction. The other lines show predictions for some representative
  values of the NP parameters.  For example, the blue curve in the
  low-$q^2$ and high-$q^2$ regions corresponds to $ (1.33 e^{-i 2.96},
  0.78 e^{i 2.47}, 0.83 e^{-i 0.27}, 3.15 e^{i 1.75})$ and $(2.15
  e^{-i 2.77},0.7 e^{-i 2.43}, 8.20 e^{-i 0.16}, 4.8 e^{-i 1.62})$,
  respectively.
\label{fig:imAT1-va}}
}

The second TP is $A_{LT}^{(im)\bar{B}(B)}$, introduced above in
Eq.~(\ref{defn-ALT}). In terms of the coupling constants and matrix
elements defined in Ref.~\cite{Alok:2010zd}, $A_{LT}^{(im)\bar{B}(B)}$
can be written as
\bea
 \label{ALT1-expr2}
A^{(im)\bar{B}}_{LT} &\!=\!& \frac{I^{LT}_4}{4 (A^{\bar{B}}_L + A^{\bar{B}}_T)} ~,\quad
A^{(im) B }_{LT} =  \frac{\bar{I}^{LT}_4}{4 (A^B_L + A^B_T)}~.
\eea
We observe that $A_{LT}$ depends on the VA couplings, as well as on
V-S and SP-T interference terms. The CP-violating triple-product
asymmetry is
\bea
\label{ALTimdef}
A_{LT}^{(im)} &=& \frac{1}{2} (A_{LT}^{(im)\bar{B}}- A_{LT}^{(im)B}) ~.
\eea

\begin{figure}[t]
\begin{center}
\includegraphics[width=0.4\linewidth]{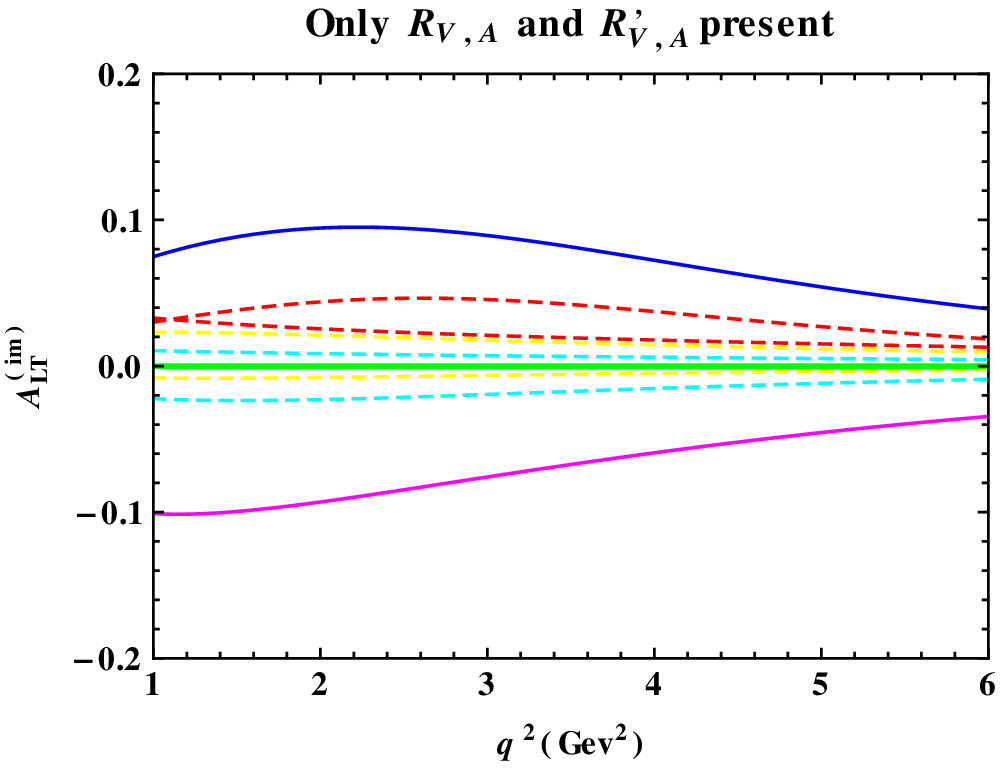}
\caption{The figure shows $A^{(im)}_{LT}(q^2)$ for $\BKstarmumu$ in
  the low-$q^2$ region, in the scenario where $(R_V, R_A, R^\prime_V,
  R^\prime_A)$ terms are all present.  The green line corresponds to
  the SM prediction.  The other lines show predictions for some
  representative values of the NP parameters.  For example, the blue
  curve corresponds to $ (1.68 e^{i 1.92}, 2.27 e^{i 0.53}, 4.22 e^{i
    0.28}, 0.14 e^{-i 1.91})$.
\label{fig:imALT-va}}
\end{center}
\end{figure}

Fig.~\ref{fig:imALT-va} shows $A^{(im)}_{LT}(q^2)$ for $\BKstarmumu$
in the presence of new VA couplings.  We make the following
observations:
\begin{itemize}

\item If only $R_{V,A}$ couplings are present, $A^{(im)}_{LT}(q^2)$
  can be enhanced up to 6\% at very low $q^2$, and is almost same as
  the SM prediction ($\approx 0 $) at high $q^2$. It can have either
  sign at both low and high $q^2$.

\item If only $R'_{V,A}$ couplings are present, $A^{(im)}_{LT}(q^2)$
  can be enhanced up to 8\% at low $q^2$ and is almost same as the SM
  prediction ($\approx 0 $) at high $q^2$.  It can have either sign at
  both low and high $q^2$.
 
\item When both primed and unprimed VA couplings are present,
  $A^{(im)}_{LT}(q^2)$ can be enhanced up to 10\% at low $q^2$ and up
  to 0.5\% at high $q^2$. It can have either sign at both low and high
  $q^2$ (see Fig.~\ref{fig:imALT-va}).

\end{itemize}

Note that $A_{LT}^{(im)}(q^2)$ is related to the observable $A_7^D$ in
Ref.~\cite{Bobeth:2008ij}: $A_{LT}^{(im)}(q^2) \approx A_7^D/8$.  Our
limit of 10\% on the maximum value of $A_{LT}^{(im)}(q^2)$ is then
consistent with the limit of 0.76 on the average value $\langle A_7^D
\rangle$ over the low-$q^2$ region, as calculated in
Ref.~\cite{Bobeth:2008ij}. However, in addition we present the full
$q^2$-dependence of this quantity.

In principle, $A_{LT}^{(im)\bar{B}(B)}$ can be generated due to NP
SP-T interference. However, we find that the effect is tiny:
$A^{(im)}_{LT}(q^2)$ can be enhanced up to 0.4\% at low $q^2$ and can
have either sign; $A^{(im)}_{LT}(q^2)$ is same as the SM ($\simeq 0$)
at high $q^2$.

\section{Discussion and summary}
\label{summary}

Even after the successful start of the LHC that will search for new
physics (NP) at the TeV scale and beyond, $B$ decays still remain one
of the best avenues of detecting indirect NP signals.  The copious
amount of data on $B$ decays, expected from future experiments like
the LHC and super-$B$ factories, will allow us to explore in detail
many decay modes that are currently considered to be rare.  The
combined analysis of many such decay modes will allow us to look for
NP in a model-independent manner.

We consider all possible Lorentz structures of new physics (NP) in the
$b \to s \mu^+ \mu^-$ transition, and analyze their effects on the
CP-violating observables in (i) $\Bsmumu$, (ii) $\BXsmumu$, (iii)
$\Bsmumugamma$, (iv) $\BKmumu$, (v) $\BKstarmumu$, and their
CP-conjugate modes.  These are the same modes we explored in the
companion paper \cite{Alok:2010zd}, where we considered only
CP-conserving quantities.  We find that for $\Bsmumu$, the only
CP-violating quantities that can be constructed even in principle
require the measurement of muon polarization, a task not possible in
foreseeable detectors. Therefore, we do not dwell on this mode
further.  For the rest of the modes, we focus on
\begin{itemize}

\item CP violation in the differential branching ratio ($A_{CP}$), and 

\item CP violation in the forward-backward asymmetry ($\Delta A_{FB}$).

\end{itemize} 
In addition, for $\BKstarmumu$, we analyze
\begin{itemize} 

\item the CP asymmetry in the longitudinal polarization fraction ($\Delta f_L$),

\item the CP asymmetries $\Delta A_T^{(2)}$ and $\Delta A_{LT}$ arising
in the angular distributions, and

\item the triple-product (TP) CP asymmetries $\Delta A_T^{(im)}$ and
$\Delta A_{LT}^{(im)}$.

\end{itemize}

We determine the constraints on the coupling constants in the
effective NP operators by using the currently available data.  On the
basis of these limits and general arguments, we expect that the
CP-violating quantities in most of the modes can only be sensitive to
the vector-axial vector (VA) couplings, while the scalar-pseudoscalar
(SP) and the tensor (T) NP operators can only contribute, if at all,
to certain TP asymmetries.  Our later detailed exploration of the
allowed parameter space for all the NP couplings vindicates this
argument.  The effects of SP and T NP operators are therefore
discussed only briefly in this paper.

\afterpage{\clearpage}

\TABLE[h!]{
{\small
\begin{tabular}{p{2.6cm}|p{2.5cm}|p{2.7cm}|p{2.6cm}|p{2.6cm}}
\hline
Observable & SM & {Only new VA} & {Only new SP} & {Only new T} \\
\hline
$\BXsmumu$ & & & & \\
\hfill $A_{\rm CP}$ & $\bullet$ $10^{-3} \to 10^{-4}$ \newline
(low$\to$high $q^2$) &
$\bullet$ $(6\to 12)\%$ \newline
(low$\to$high $q^2$)
& $\bullet$ Marginal S & $\bullet$ Marginal S/E \\
\hfill $\Delta A_{FB}$ & $10^{-4} \to 10^{-5}$ \newline
(low$\to$high $q^2$)
& $\bullet$ $(3 \to 12)\%$ \newline
(low$\to$high $q^2$)
& $\bullet$ $<$ 1\% &  No effect\\
 & & & & \\
\hline
$\Bsmumugamma$ & & & & \\
\hfill $A_{\rm CP}$ & $\bullet$ $10^{-3} \to 10^{-4}$ \newline
(low$\to$high $q^2$) &
$\bullet$ $(30 \to 13)\%$ \newline
(low$\to$high $q^2$)
&  No effect & $\bullet$ $<$ 1\% \\
\hfill $\Delta A_{FB}$ & $10^{-4} \to 10^{-5}$ \newline
(low$\to$high $q^2$)
& $\bullet$ $(40 \to 18)\%$ \newline
(low$\to$high $q^2$)
& No effect  & $\bullet$ $<$ 1\%\\
 & & & & \\
\hline
$\BKmumu$ & & & & \\
\hfill $A_{\rm CP}$ & $\bullet$ $10^{-3} \to 10^{-4}$ \newline
(low$\to$high $q^2$) &
$\bullet$ $(7 \to 12)\%$ \newline
(low$\to$high $q^2$)
& $\bullet$ Marginal S & $\bullet$ Marginal S/E \\
\hfill $\Delta A_{FB}$ & Zero
&  No effect
& $\bullet$ $<$ 1\%  & No effect\\
 & & & & \\
\hline
$\BKstarmumu$ & & & & \\
\hfill $A_{\rm CP}$ &$\bullet$ $10^{-3} \to 10^{-4}$ \newline
(low$\to$high $q^2$) & $\bullet$ $(9 \to 14)\%$ \newline
(low$\to$high $q^2$) &  No effect &$\bullet$ $<$ 1\%   \\
 & & & & \\
\hfill $\Delta A_{FB}$ &$\bullet$ $10^{-4} \to 10^{-6}$ \newline
(low$\to$high $q^2$) & $\bullet$ $(6 \to 19)\%$ \newline
(low$\to$high $q^2$) & No effect &   $\bullet$ $<$ 1\%     \\
 & & & & \\
\hfill $\Delta f_L$ & $\bullet$ $10^{-4} \to 10^{-7}$ \newline
(low$\to$high $q^2$) & $\bullet$ $(9 \to 16)\%$ \newline
(low$\to$high $q^2$) & No effect  &  $\bullet$ $<$ 1\%    \\
 & & & & \\
\hfill $\Delta A_T^{(2)}$ & Zero & $\bullet$ $\sim 12\%$ &  No effect &  No effect    \\
 & & & & \\
\hfill $\Delta A_{LT}$ & Zero& $\bullet$ $< 3\%$ & No effect &  No effect    \\
 & & & & \\
 \hfill $A^{(im)}_{T}$ &Zero &$\bullet$ $\sim 50\%$  & No effect &
No effect   \\
 & & & & \\
  \hfill $A^{(im)}_{LT}$ & Zero & $\bullet$ $\sim 10\%$ & No effect &  No effect   \\
 && & & \\
\hline
\end{tabular}
}
\label{tab:summary}
\caption{The effect of NP couplings on observables.
E: enhancement, S: suppression. 
The numbers given are optimistic estimates.}
}

On the other hand, the VA operators can have a significant impact on
the CP-violating observables. (See Table~\ref{tab:summary}).  The SM
predicts $A_{CP}(q^2) \lesssim 10^{-3}$ for all the modes, while VA NP
operators allow this quantity to be as large as $\sim 10\%$ (for
$\BXsmumu, \BKmumu$ and $\BKstarmumu$) and even up to $\sim 30 \%$ for
$\Bsmumugamma$.  Even $\Delta A_{FB}$, expected to be $\lesssim
10^{-4}$ in the SM, can be enhanced up to $\sim 10\%$ (for $\BXsmumu$)
and up to $\sim 40\%$ (for $\Bsmumugamma$).  While $\Delta A_{FB}$ in
$\BKmumu$ stays zero even with VA NP, its value in $\BKstarmumu$ may
be enhanced to $\sim 10\%$ from its SM expectation of $\lesssim
10^{-4}$.

In $\BKstarmumu$ the SM predicts $\Delta f_L \lesssim 10^{-4}$, while
VA NP operators allow this quantity to be enhanced up to $\sim 10\%$.
$\Delta A^{(2)}_T$, $\Delta A_{LT}$, $ A^{(im)}_T$ and $
A^{(im)}_{LT}$ are all zero in the SM. VA NP operators can enhance
$\Delta A^{(2)}_T$ up to $\sim 12\%$, $ A^{(im)}_T$ even up to $\sim
50\%$, and $ A^{(im)}_{LT}$ up to $\sim 10\%$. $\Delta A_{LT}$ can not
be enhanced more than $\sim 3\%$ even in the presence of VA NP
operators.  Note that while in almost all the cases the impact of the
left-handed VA NP couplings $R_{V,A}$ is dominant, for the TP
asymmetry $\Delta A_T^{(im)}$, the $R'_{V,A}$ couplings play a
dominating role.

TP's can also be generated by NP-NP interference. However, we do not
find large effects. The interference of SP-T operators can increase
$A^{(im)}_{LT}(q^2)$ up to only 0.4\% at low $q^2$.

It is quite possible that if the NP is of the VA type, its presence
would first be indicated through the CP-conserving/CP-averaged
quantities considered in Ref.~\cite{Alok:2010zd}.  However, the
CP-violating signals considered in this paper are so robust (orders of
magnitude more than the SM predictions) that these may be the ones
that will unambiguously establish the presence of NP of the VA kind.
Moreover, hadronic uncertainties play a very minor role in the
CP-violating asymmetries considered in this paper.  A combined
analysis of CP-violating and CP-conserving signals may allow even the
determination of the magnitudes and phases of the NP coupling
constants, in addition to confirming the NP Lorentz structure.


\acknowledgments 

We thank S. Uma Sankar for helpful collaboration on several parts of
this analysis.  The work of AKA and DL was financially supported by
NSERC of Canada.  The work of A. Datta and M. Duraisamy was supported
by the US-Egypt Joint Board on Scientic and Technological Co-operation
award (Project ID: 1855) administered by the US Department of
Agriculture.


\end{document}